\documentclass[preprint]{aastex}
\usepackage{multirow}
\shorttitle{Formation of planetesimals}
\shortauthors{Michikoshi et al.}

\begin{document}
\title{$N$-Body Simulation of Planetesimal Formation through Gravitational Instability of a Dust Layer in Laminar Gas Disk}

\author{Shugo Michikoshi\altaffilmark{1}, 
Eiichiro Kokubo\altaffilmark{1,2}, and
Shu-ichiro Inutsuka\altaffilmark{3} 
}
\altaffiltext{1}{
Center for Computational Astrophysics, National Astronomical Observatory of Japan, Osawa, Mitaka, Tokyo 181-8588, Japan
}
\altaffiltext{2}{
Division of Theoretical Astronomy, National Astronomical Observatory of Japan, Osawa, Mitaka, Tokyo 181-8588, Japan
}
\altaffiltext{3}{
Department of Physics, Graduate School of Science, Nagoya University, Furo-cho, Chikusa-ku, Nagoya, Aichi 464-8602, Japan
}
\email{michikoshi@cfca.jp, kokubo@th.nao.ac.jp, and inutsuka@nagoya-u.jp}
\begin{abstract}
We investigate the formation process of planetesimals from the dust layer by the gravitational instability in the gas disk using local $N$-body simulations.   
The gas is modeled as a background laminar flow.
We study the formation process of planetesimals and its dependence on
 the strength of the gas drag. 
Our simulation results show that the formation process is divided into 
 three stages qualitatively: the formation of wake-like density
 structures, the creation of planetesimal seeds, and their collisional
 growth.   
The linear analysis of the dissipative gravitational instability shows
 that the dust layer is secularly unstable although Toomre's $Q$ value
 is larger than unity. 
However, in the initial stage, the growth time of the gravitational
 instability is longer than that of the dust sedimentation and the
 decrease in the velocity dispersion. 
Thus, the velocity dispersion decreases and the disk shrinks vertically.
As the velocity dispersion becomes sufficiently small, the
 gravitational instability finally becomes dominant.  
Then wake-like density structures are formed by the gravitational
 instability.   
These structures fragment into planetesimal seeds. 
The seeds grow rapidly owing to mutual collisions. 
\end{abstract}

\section{Introduction \label{sec:intro}}
In the standard scenario of planet formation, planetesimals are
the precursors of planets.
Their formation process is one of the unsolved problems of the planet formation theory. 
Beginning with micron-sized dust grains, they grow to centimeter size in
 a protoplanetary disk via collisional agglomeration
 \citep{Weidenschilling1980,Blum2000}.
The least understood growth phase is that of growth from centimeter size to
 kilometer size. 
Since the gas drag of meter-sized aggregates is weak, they are slightly
 decoupled from gas with the sub-Keplerian velocity.
The resulting head-wind decreases their angular momentum and causes
 inward drift, whose timescale is about a few hundred years
 \citep{Adachi1976, Weidenschilling1977}. 
The growth process in this stage must be faster than this drift
 timescale.   
Another problem is the sticking probability of the aggregates.
In this stage, since the threshold velocity for the destruction is low,
 their collisions may lead to destruction
 \citep[e.g.,][]{Blum2000,Sirono2004}.  

A very thin and dense layer of settled dust aggregates in the mid-plane
 of the protoplanetary disk is gravitationally unstable.
Then the gravitational collapse of the dust layer occurs, and
 kilometer-sized planetesimals are formed directly \citep{Safronov1969,
 Goldreich1973}.
This scenario has the advantage of a very rapid formation timescale, which
 is on the order of the Keplerian time, thus avoiding the migration of
 meter-sized aggregates. 
However, turbulence in the protoplanetary disk may prevent dust from
 settling to the mid-plane. 
As dust settles in the protoplanetary disk, a vertical shear velocity
 develops because the dust-rich gas in the mid-plane rotates with the
 velocity closer to the Keplerian velocity than that of the
 dust-depleted gas, which may cause turbulence.
The turbulence can mix dust with gas and prevent the dust layer from
 settling into a dense enough sheet for gravitational instability
 \citep{Weidenschilling1980}.   
Problems concerning the gravitational instability of the dust layer still remain unsolved. 

There are two methods for calculating the dynamics of dust particles in
 gas: fluid simulation and $N$-body simulation. 
One can consider the particles as virtually fluid and treat them with the Euler equation or Navier-Stokes equation. 
For example, \citet{Yamoto2006} performed the two-dimensional numerical
 simulation to investigate the density evolution of the dust layer due
 to gravitational instability.
They assumed that the dust layer is axisymmetric with respect to the
 rotational axis. 
They treated the dust component as a pressure-free fluid because the
 velocity dispersion of dust is negligible if the stopping time
 $t_\mathrm{stop}$ is sufficiently short, where the stopping time
 $t_\mathrm{stop}$ is the characteristic timescale for a particle to
 stop in gas. 
They found that the dust layer becomes extremely thin if the stopping time
 is long.
\citet{Wakita2008} performed the numerical simulation of the
 gravitational instability of a two-dimensional thin disk and
 investigated the non-axisymmetric modes.

On the other hand, we can investigate the dynamics of particles in gas by using $N$-body simulations \citep{Tanga2004, Michikoshi2007, Michikoshi2009, Rein2010}. 
This is straightforward and precise, but the calculation is very time-consuming and the practical number of particles is limited. 
\cite{Tanga2004} performed $N$-body simulations of a gravitationally
unstable disk with a local-shearing box.  They investigated the formation of
clumps of planetesimals at t30 AU by the gravitational instability.  
They considered the drag force from the background
gas.  They showed that the planetesimal clumps form owing to the
gravitational instability, which correspond to planetesimals in our
calculation.  In their simulation, the optical depth is smaller than that in
our calculation and the particle disk is unstable even initially.  We consider
the formation of planetesimals at 1 AU, thus we use the large optical depth, and
assume the dust layer is initially stable.
\cite{Michikoshi2007} performed a set of local simulations of the
 self-gravitating collisional particle disks without gas using $N$-body
 simulation (hereafter Paper I).     
Dust particle dynamics is calculated with the Hill equations in a
 local-shearing box \citep{Wisdom1988}. 
They adopted the rubble pile model (hard and soft sphere models) as
 collision models. 
They found that the formation process is divided into three stages: the 
 formation of non-axisymmetric wake-like structures \citep{Salo1995},
 the creation of aggregates, and the rapid collisional growth of the
 aggregates. 
The mass of the largest aggregates is larger than the mass predicted by
 the linear perturbation theory.
\cite{Michikoshi2009} adopted the alternative model of collisions,
 `accretion model' (Paper II). 
In the accretion model, the number of particles decreases as the
 calculation proceeds; thereby this model enables us to perform
 large-scale and long-term simulations. 
They obtained the final mass of planetesimals as a function of the size
 of the computational box. 
\cite{Rein2010} performed the numerical simulations to investigate the validity of the super-particle approximation.
They considered the various physics, such as the gas drag, the self-gravity, the physical collision and the turbulence.
They treated the collisions as the hard sphere model.
They investigated the numerical requirements to study the gravitational instability.

To understand the basic dynamics of gravitational instability, we neglected the effect of gas in Papers I and II.
However, it is obvious that the gas plays an important role in planetesimal formation when dust particles are small. 
The interaction with gas through drag is dominant in the dynamical evolution of particles. 
Many authors investigated the effect of gas.
The particles drift radially due to gas drag \citep{Adachi1976,
Weidenschilling1977, Nakagawa1986}.  As the sedimentation of dust aggregates
towards the midplane proceeds, the vertical velocity shear increases.  Thus,
the Kelvin-Helmholtz instability occurs \citep{Weidenschilling1980, Cuzzi1993,
Sekiya1998, Dobrovolskis1999, Sekiya2000, Sekiya2001, Ishitsu2002, Ishitsu2003,
Gomez2005, Michikoshi2006, Johansen2006, Chiang2008, Barranco2009}.  The
Kelvin-Helmholtz instability makes the dust layer turbulent.  
The turbulent gas prevents the concentration of dust \citep{Weidenschilling1993}, and helps the concentration \citep{Barge1995,Fromang2006}.  
The interaction between gas and dust causes the streaming instability \citep{Youdin2005}.  
By the streaming instability, the dust grains concentrate strongly and the large Ceres-sized planetesimals form \citep{Youdin2007a, Johansen2007b, Johansen2007, Johansen2009}.  
The loss of angular momentum due to the gas drag helps gravitational instability \citep{Ward1976, Youdin2005a, Youdin2005b}.  
The time-scale for the dissipative instability is relatively slow, but the dust layer may be unstable even when
the Toomre's Q value is larger than unity.  
As a first step toward understanding the effect of gas on gravitational instability, we introduce gas as a background laminar
 flow in the present paper.
The aim of the present paper is to investigate the effect of the laminar 
 gas on the planetesimal formation through gravitational instability.
The laminar flow causes the dissipation of the kinetic energy and thus
 radial migration of dust.   

In \S 2, we summarize the results of the linear analysis of the
 gravitational instability under gas drag. 
In \S 3, we describe the simulation method and the initial condition.
In \S 4, we present our numerical results. 
\S 5 is devoted to discussions.
We summarize the results in \S 6.

\section{Dispersion Relation for Gravitational Instability under Gas
 Drag \label{sec:disprel}}

The linear stability analyses of the dust layer with a finite thickness
 were performed by \cite{Sekiya1983} and \cite{Yamoto2004}. 
They assumed that the size of dust aggregates is sufficiently small, in
 other words, dust is fully coupled with gas, thus they treated gas and
 dust as one fluid.  
That introduces no dissipation due to gas drag.
They considered the vertical structure of the dust layer, and studied
 the Roche criterion.
The linear stability analyses of dissipative gravitational instability
 were performed by \cite{Ward1976}, \cite{Coradini1981}, and
 \cite{Youdin2005a}.   
In this section, we summarize the essence of their results.
\cite{Ward1976} and \cite{Youdin2005a} used one component model, i.e.,
 they neglected the dynamics of gas flow, but they consider the gas drag from the stationary background gas flow. 
Using their results, we discuss conditions for the gravitational instability under the influence of gas drag.

\subsection{Dispersion Relation \label{sec:dispersion}}

We adopt the isothermal equation of state for the dust layer:
\begin{equation}
  p = c^2 \rho,
\end{equation}
 where $p$ is the pressure, $c$ is the isothermal
 sound speed, and $\rho$ is the density of the dust layer. 
In this equation of state, the velocity dispersion of dust is always
 constant. 
Clearly, the velocity dispersion must not be constant.
It changes owing to several processes such as gas drag, inelastic
 collisions, and gravitational scattering. 
This assumption gives us the simple analytical expression of the
 dispersion relation, with which we can grasp the nature of dissipative
 gravitational instability. 

We assume that the gas velocity is equal to the Kepler velocity.
We consider the reference point that is at the semi-major axis $a_0$.
The Kepler angular velocity of the reference point is
 $\Omega = \sqrt{GM_\mathrm{s}/a_0^3}$ where $M_\mathrm{s}$ is the mass
 of the central star and $G$ is the gravitational constant.
We introduce the local Cartesian coordinates in which the $x$-axis is directed radially, the $y$-axis  follows the direction of rotation,
 and the $z$-axis follows the direction perpendicular to mid-plane.   
We define
 $\mathbf v = (v_{x}, v_{y},
 v_{z})$ as the deviation velocity field from the local Kepler
 velocity. 

The basic equations for the dust layer in this frame are given by 
\begin{equation}
  \frac{\partial \rho}{\partial t} + \nabla\cdot (\rho \mathbf v) = \frac{3}{2} \Omega x \frac{\partial \rho}{\partial y}, 
\end{equation}
\begin{equation}
  \frac{\partial \mathbf v}{\partial t} + (\mathbf v \cdot \nabla) \mathbf v =-\frac{\nabla p}{\rho} + (2v_{y}\Omega, -\frac{1}{2}v_{x} \Omega, -z\Omega^2)+\frac{3}{2} \Omega x \frac{\partial \mathbf v}{\partial y} - \frac{1}{t_\mathrm{stop}}\mathbf v-\nabla \phi, 
\end{equation}
\begin{equation}
  \nabla^2 \phi = 4 \pi G \rho,
\end{equation}
 where $\phi$ is the gravitational potential due to the
 self-gravity.
The simulation results in \S 4 show that the density fluctuation is not
 axisymmetric. 
However, for the sake of simplicity, restricting ourselves to the
 axisymmetric mode ($\partial / \partial y = 0$), we carry out a normal
 mode analysis in the form of $\exp(-i(\omega t - kx))$ where $k$ is the 
 wave number in the $x$-direction and $\omega$ is the growth rate.

Integrating toward $z$, we obtain the equation of perturbed quantities $\Sigma_{1},
 v_{x1}, v_{y1}, \phi_1$: 
\begin{equation}
  - i \omega \Sigma_{1} + i k \Sigma_{0}
    v_{x1}= 0, 
\end{equation}
\begin{equation}
  - i \omega v_{x1} = 2 \Omega v_{y1} - i k c^2 \frac{\Sigma_1}{\Sigma_0}+ \frac{1}{t_\mathrm{stop}} v_{x1} - i k \phi_1, 
\end{equation}
\begin{equation}
  - i \omega v_{y1} = -\frac{1}{2} \Omega
    v_{x1}-\frac{1}{t_\mathrm{stop}} v_{y1}, 
\end{equation}
\begin{equation}
  - 2 \pi G \Sigma_{1} = k \phi_{1},
\end{equation}
 where $\Sigma_{}$ is the surface density of the dust layer. 
\cite{Youdin2005a} considered the finite thickness model using the
 softening term \citep{Vandervoort1970}.
Poisson equation with the softening parameter is 
\begin{equation}
  - 2 \pi G \Sigma_{1} T(kh) = k \phi_{1},
\end{equation}
 where $h$ is the scale height of the dust layer and $T(kh)$ is the
 softening parameter: 
\begin{equation}
T(kh)=\frac{1}{1+kh}.
\end{equation}

For nontrivial solutions, the determinant of the coefficient matrix must vanish. 
Thus the following dispersion relation is obtained \citep{Youdin2005a}:
\begin{equation}
  \mu^3 + \frac{2}{t_\mathrm{stop}} \mu^2 +
  \left(\Omega^2+\frac{1}{t_\mathrm{stop}^2}-2 \pi k G \Sigma_{0} T(kh) + c^2 k^2 \right) \mu +  \frac{ c^2 k^2 - 2 \pi k \Sigma_{0} G }{t_\mathrm{stop}}
  = 0, 
  \label{eq:disp1}
\end{equation}
 where $\mu = - i \omega$. 
The mode for $\Re{(\mu)}>0$ is unstable, where $\Re{(X)}$ is the real
 part of a complex number $X$. 
In the gas-free limit ($t_\mathrm{stop} \to \infty$) and the thin disk
 limit ($kh \to 0$), the dispersion relation converges to the
 conventional dispersion relation of a thin disk
 \citep{Toomre1964,Goldreich1973}. 
Equation (\ref{eq:disp1}) can be written in a non-dimensional form
 scaled by the length $c/\Omega$ and the time $\Omega^{-1}$: 
\begin{equation}
  \tilde \mu^3 + \frac{2}{\tilde t_\mathrm{stop}}\tilde \mu^2 + \left(
  \frac{1}{\tilde t_\mathrm{stop}^2} + \tilde k^2+1 - \frac{2\tilde
  k}{Q} T(\tilde k \tilde h) \right)\tilde \mu +\frac{\tilde k^2Q -
  2\tilde k}{\tilde t_\mathrm{stop}Q}=0, 
  \label{eq:disp2}
\end{equation}
 where we used Toomre's $Q$ given by
 $Q=\Omega c /(\pi G \Sigma_{0})$.
A tilde denotes non-dimensional quantity hereafter. 

A dissipative dust layer is secularly unstable for long-wavelength modes
 although $Q>1$. 
The physical mechanism of the instability is explained as follows \citep{Goodman2000, Chiang2010}.  
We consider a thin axisymmetric overdense ring and the thick back-ground gas.  
From the force balance of the self-gravity, we find that the dust rotates faster at the
outer edge of the ring. Thus, the drag force at the outer edge is strong. This
causes the inward drift.  Conversely, the dust at the inner edge drifts
outward. These effects shrink ring radially.

\begin{figure}
 \begin{center}
  \plotone{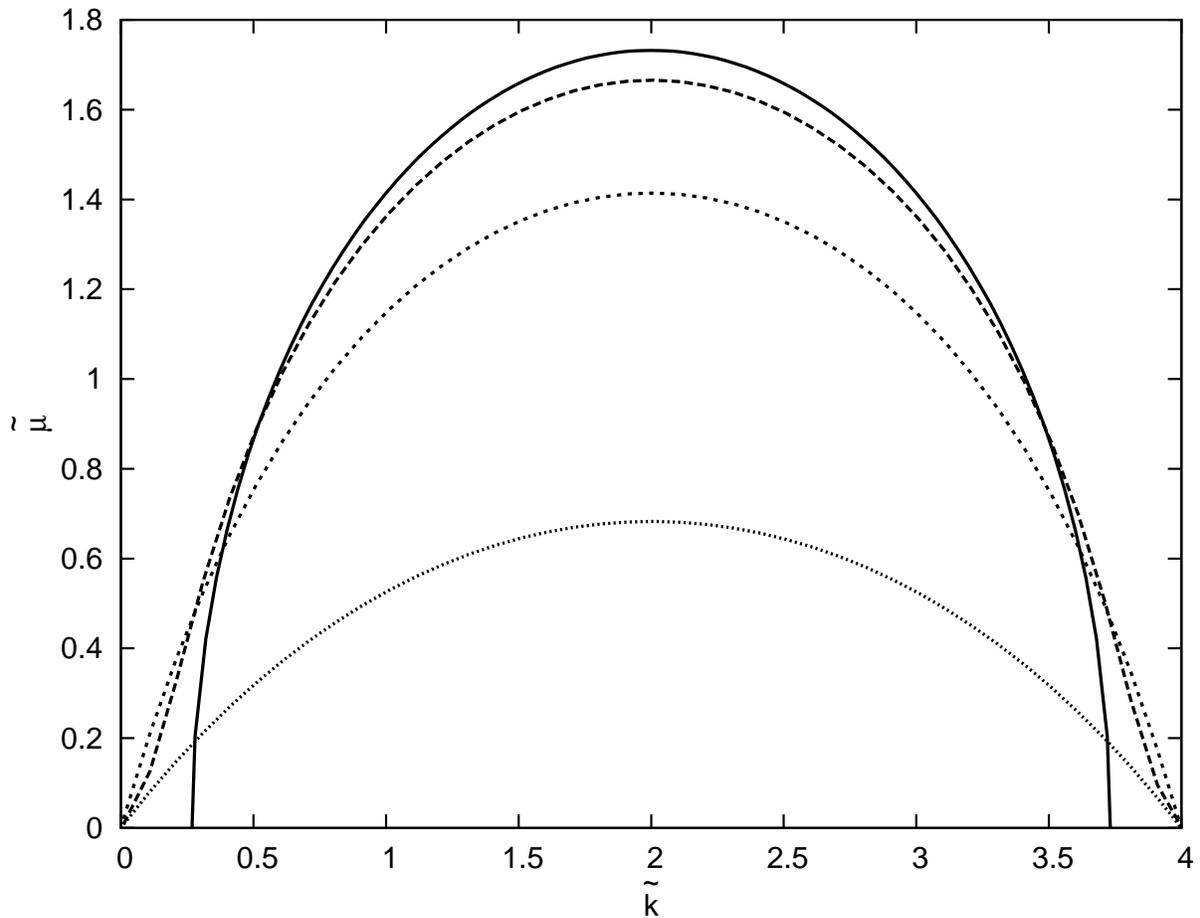}
 \end{center}
 \caption{
 The dispersion relation of the dust layer under gas drag for $Q = 0.5$.  We used the thin disk model ($h=0$). 
 The stopping time is $0.2$ (dotted curve), $1.0$ (short-dashed curve), and $2.0$ (dashed curve).  
 The solid curve denotes the dispersion relation for the gas-free model.} 
  \label{fig:disper.eps}
 \end{figure}

Figure \ref{fig:disper.eps} shows the dispersion relation of the dust
 layer under gas drag for $Q = 0.5$.  
We used the thin disk model ($h=0$). 
The stopping time is $0.2$, $1.0$, $5.0$, and $ \infty$ (gas-free).
All models have the maximum growth rate at $\tilde k = 2$.
We will see that the growth rate has the maximum value at 
 $\tilde k = 1/Q$ in \S \ref{sec:mugr}. 
As the stopping time lengthens, the growth rate shrinks.
In the gas-free model, long-wavelength modes are stable, while in the
 gas model, they are unstable.

\subsection{Growth Rate of the Most Unstable Mode\label{sec:mugr}}

As discussed in \S \ref{sec:dispersion}, the dust layer is always
 unstable for long-wavelength modes. 
However, the timescale of the instability can be longer than the
 dynamical timescale, because this is the secular instability. 
Therefore, in order to understand what happens, we have to compare the
 timescale of the instability with those of other processes such as the
 sedimentation, the radial migration, and the growth of dust
 aggregates. 

Here, we assume the thin disk approximation ($h=0$) in order to obtain
 the simple analytical formula. 
For $\tilde k<2/Q$, $\tilde \mu$ is a real positive value.
The growth rate for the most unstable mode $\tilde \mu$ is a function of
 $\tilde k$, $\tilde t_\mathrm{stop}$, and $Q$. 
From Equation (\ref{eq:disp2}), the partial derivative of $\tilde \mu$
 with respect to $\tilde k$ is given as 
\begin{equation}
  \frac{\partial \tilde \mu}{\partial \tilde k} = \frac{2 (1/Q-\tilde
  k)(\tilde \mu + 1)}{4 \tilde \mu/\tilde t_\mathrm{stop} + 1/\tilde
  t_\mathrm{stop}^2 + \tilde k^2+1 -2 \tilde k / Q + 3\tilde \mu^2}. 
  \label{eq:disp3}
 \end{equation}
Using Equation (\ref{eq:disp2}), we can show that the denominator on the
 right hand side of Equation (\ref{eq:disp3}) is positive when
 $\tilde k < 2/Q$.  
Hence, the $\tilde \mu$ has the maximum value at $\tilde k=1/Q$ for a fixed $\tilde t_\mathrm{stop}$ and $Q$.
The substitution of $\tilde k=1/Q$ into Equation (\ref{eq:disp2})
 provides the equation of the maximum growth rate $\tilde
 \mu_\mathrm{max}$: 
\begin{equation}
  \tilde \mu^3_\mathrm{max} + \frac{2}{\tilde t_\mathrm{stop}}\tilde
  \mu^2_\mathrm{max} + \left( \frac{1}{\tilde t_\mathrm{stop}^2} +1 -
  \frac{1}{Q^2} \right)\tilde \mu_\mathrm{max} -\frac{1}{\tilde
  t_\mathrm{stop}Q^2}=0. 
\label{eq:mumax}
\end{equation}

To obtain the condition that the gravitational instability dominates over the other process (symbolically denoted by ``x''), we equate $\tilde T_X$ with $\tilde \mu_\mathrm{max}^{-1}$ where $T_X$ is the characteristic timescale of a process X, ans solve the resultant equation with respect to Q to obtain 
 $Q$ value:  
\begin{equation}
  Q_\mathrm{crit,X} = \tilde T_X \sqrt{ \frac{\tilde
   t_\mathrm{stop}(\tilde T_X + \tilde t_\mathrm{stop})}{\tilde
   t_\mathrm{stop}^2\tilde T_X^2+(\tilde T_X+\tilde
   t_\mathrm{stop})^2}}. 
\label{eq:crit_Q}
\end{equation}
When $Q<Q_\mathrm{crit,X}$, gravitational instability is a dominant process compared to the process X. 
On the other hand, when $Q>Q_\mathrm{crit,X}$, the timescale for the
 process X is shorter than the timescale of gravitational
 instability. 
Although the dust layer is unstable, the process X is dominant. 

We can derive the leading term of the power series expansion with respect to
$t_\mathrm{stop}$ for Equation (13) and find the asymptotic solution
\citep{Ward1976, Youdin2005a, Chiang2010}.  When the gas drag is strong $\tilde
t_\mathrm{stop} \ll 1$, the growth rate for the most unstable mode is 
\begin{equation}
\tilde \mu_\mathrm{max} = \frac{\tilde t_\mathrm{stop}}{Q^2}.
\end{equation}
Conversely, when the gas drag force is weak $t_\mathrm{stop} \gg 1$ and $Q>1$,
the growth rate for the most unstable mode is 
\begin{equation}
\tilde \mu_\mathrm{max} = \frac{1}{(Q^2-1) \tilde t_\mathrm{stop}}.
\end{equation}
When $Q<1$, the growth rate for the most unstable mode converges to the
finite value $\sqrt{1/Q^2-1}$ as $\tilde t_\mathrm{stop} \to \infty$, which
is the maximum growth rate for the gas-free gravitational instability.

\section{Method of Calculation }

\subsection{Model}

The method of calculation is the same as those used in Papers I and II
 except for including gas,  
 which is based on the method of the local simulation of planetary rings
 \citep[e.g.,][]{Wisdom1988,Richardson1994,Daisaka1999}. 

We describe quantities in the non-dimensional form independent of the
 semi-major axis $a_0$, and the mass of the central star $M_\mathrm{s}$
 by scaling the time by $\Omega_0^{-1}$, the length by the Hill radius
 $\tilde h a_0 = r_\mathrm{H}$, and the mass by $\tilde h^3M_\mathrm{s}$, where
 $\tilde h = (2m_\mathrm{p}/3M_\mathrm{s})^{1/3}$ and $m_\mathrm{p}$ is the
 initial mass of particles \citep{Petit1986, Nakazawa1988}. 
The initial mass of particles $m_\mathrm{p}$ is assumed to be
 identical. 
The non-dimensional Hill equations for the particle $i$ are given by
 \citep[e.g.,][]{Nakazawa1988}: 
\begin{eqnarray}
	\frac{d^2 \tilde x_i}{d \tilde t^2} &=&  2\frac{d \tilde y_i}{d \tilde t} +3 x_i  
		+ \sum_{j} \frac{\tilde m_j}{\tilde r_{ij}^3}(\tilde x_j-\tilde x_i)
  		-\frac{1}{\tilde t_{\mathrm{stop},i}}\left(\frac{d \tilde x_i}{d \tilde t}-\tilde v_{\mathrm{g}x}\right), \label{eq:Hilleqx} \\
	\frac{d^2 \tilde  y_i}{d \tilde t^2} &=& -2\frac{d \tilde x_i}{d \tilde t} 
		+ \sum_{j} \frac{\tilde m_j}{\tilde r_{ij}^3}(\tilde y_j-\tilde y_i)
  		-\frac{1}{\tilde t_{\mathrm{stop},i}}\left(\frac{d \tilde y_i}{d \tilde t}-\tilde v_{\mathrm{g}y}\right), \label{eq:Hilleqy} \\
	\frac{d^2 \tilde z_i}{d \tilde t^2} &=& -\tilde z_i 
		+ \sum_{j} \frac{\tilde m_j}{\tilde r_{ij}^3}(\tilde z_j-\tilde z_i)
  		-\frac{1}{\tilde t_{\mathrm{stop},i}}\left(\frac{d \tilde z_i}{d \tilde t}-\tilde v_{\mathrm{g}z}\right), \label{eq:Hilleqz}
\end{eqnarray}
 where $(\tilde x_i,\tilde y_i,\tilde z_i)$, $\tilde m_i$, and
 $\tilde t_{\mathrm{stop},i}$ are the position, mass, and the stopping
 time of the particle $i$, 
 $\tilde r_{ij}$ is the distance between particles $i$ and $j$, and
 $(\tilde v_{\mathrm{g}x},\tilde v_{\mathrm{g}y},\tilde v_{\mathrm{g}z})$ 
 is the velocity field of gas.

We neglect the dust back-reaction on gas for the sake of simplicity. 
We assume that the laminar gas flow, which has the sub-Keplerian
 velocity field: 
\begin{equation}
(\tilde v_{\mathrm{g}x},\tilde v_{\mathrm{g}y},\tilde v_{\mathrm{g}z}) =
 \left(0, - \frac{3}{2} \tilde x - \tilde v_{\mathrm{dif}},0 \right), 
\end{equation}
 where $-3 \tilde x/2$ is the Keplerian shear velocity, 
 $\tilde v_{\mathrm{dif}}$ is the difference from the Keplerian
 velocity.  
The stopping time $\tilde t_{\mathrm{stop},i}$ depends on the size of
 particle: $\tilde t_{\mathrm{stop},i} = \tilde t_{\mathrm{stop},0} 
 \left( \tilde m_i/\tilde m_p \right)^{\alpha}$ where $\alpha$ is the
 power-law index, and $\tilde t_{\mathrm{stop},0}$ is the stopping time
 of the particle with $\tilde m_\mathrm{p}$.  
In this paper, we adopt $\alpha = 0$ and 2/3. 
The index $\alpha=2/3$ corresponds to Stokes' law
 \citep[e.g.,][]{Landau1959}.  

The calculation box has the periodic boundary condition
 \citep[e.g.,][]{Wisdom1988}. 
The size of the box is taken as a square
 $A \lambda_\mathrm{m} \times A \lambda_\mathrm{m}$ where
 $\lambda_\mathrm{m}$ is the most unstable wavelength of gravitational
 instability and $A$ is the non-dimensional parameter.

The system is characterized by two non-dimensional parameters, the
 optical depth and the ratio of the Hill radius to the diameter of a
 particle \citep{Daisaka1999}: 
\begin{equation}
\tau = \frac{3\Sigma}{4 \rho_\mathrm{p} r_\mathrm{p}},
\end{equation}
\begin{equation}
\zeta = \frac{r_\mathrm{H}}{2r_\mathrm{p}},
\end{equation}
 where $\rho_\mathrm{p}$ and $r_\mathrm{p}$ are the internal density and
 initial radius of the particle, respectively.
The realistic value for ``solid'' dust particles is $\zeta \simeq 105.28$ for 
 $\rho_\mathrm{p} = 2 \mathrm{g/cm^3}$ and $a_0 = 1 \mathrm{AU}$.
However, we adopt $\zeta \simeq 2$ in this paper owing to the
 computational limit (see Papers I and II).  
We set the optical depth $\tau =0.1$, which is approximately equal to the
 realistic value $\tau=0.19$ for $\Sigma_{\rm d} = 10 \mathrm{g/cm^3}$,
 $r_\mathrm{p} = 20 \mathrm{cm}$, and $\rho_\mathrm{p} = 2 \mathrm{g/cm^3}$. 

Since our choice of $\zeta$ is smaller than the realistic value, a particle used in the simulation is not a realistic dust
 particle but a `super-particle' that represents a group of many small
 particles that have the same position and velocity.
Thus the restitution coefficient $\epsilon$ corresponds to the rate of
 the dissipation due to collisions between super-particles.
The low value of $\zeta$ means that the physical size of a particle in our
simulation is very large.  This implies that the effect of the gravitational
scattering is relatively weaker than that of collisions.  We investigated the
$\zeta$ value dependence in paper I and II in the narrow range ($\zeta=1.5-3.0$)
and we confirmed that the formation process does not change in this range
qualitatively.  
However, we show that the collisional growth is important in the final stage where $|zeta$ controls the growth rate.  
Thus, we should perform
the large $\zeta$ value.  In the future work, we will concentrate on the
collisional growth and perform the $N$-body simulations with a realistic
$\zeta$ value.

In the Hill coordinates, the Keplerian orbit is determined by six
 parameters: 
 the position of the guiding center, eccentricity $e$, inclination $i$, 
 and two phases for epicyclic and vertical oscillations 
\citep{Nakazawa1988}.
The initial eccentricity and inclination of particles are assumed to
 follow the Rayleigh distribution.  
We fix $\sqrt{\langle e^2 \rangle / \langle i^2 \rangle} = 2$ according to \cite{Ida1992}.

The other parameters are uniformly distributed, avoiding overlapping.

We adopt the hard-sphere and accretion models as a collision model
 (Papers I and II).  
In the hard-sphere model, the penetration of particles is not possible 
 \citep[e.g., paper I, ][]{Richardson1994}.   
When a collision occurs, the particle velocity changes instantly.
The relative tangential velocity is conserved and the magnitude of the
 relative normal velocity is reduced by a factor of $\epsilon$.  
We use the constant $\epsilon$.
We also use the accretion model \citep[e.g., Paper II, ][]{Kokubo2000a}. 
In this model, when two particles collide, if the binding condition is
 satisfied, the two particles merge into one particle. 
The binding condition is described by the Jacobi integral $J<0$ 
 \citep[e.g., Paper II,][]{Nakazawa1988}.
	
In the hard-sphere model, since a planetesimal consists of many small particles, it can be distorted in shape and break up. 
On the other hand, in the accretion model, a planetesimal cannot be distorted or break up. 
In this respect, the hard-sphere model is more realistic than the accretion
 model. 
However the accretion model has an advantage.
First, since the number of particles decreases as the calculation
 proceeds, this model enables us to reduce the calculation cost. 
Second, we can easily handle size-dependent drag coefficient.
In general, the drag coefficient depends on the size of a particle.
In the hard sphere model, it is complicated to handle the size-dependent
 drag coefficient. 
In the accretion model, we can adopt the size-dependent drag without
 any difficulty. 
We have shown that results in the accretion model are the same as those in the
 hard sphere model in the gas-free model (Paper II).

\subsection{Test Simulations}

\begin{figure}
  \begin{center}
	\plottwo{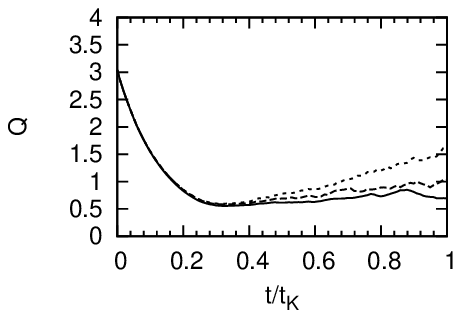}{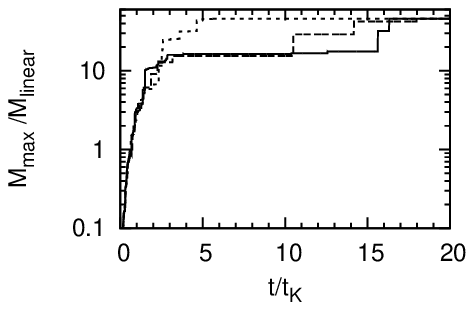}
  \end{center}
  \caption{The time evolutions of Toomre's $Q$ value (\textit{left
  panel}) and the maximum mass $M_\mathrm{max}$ of planetesimals
  (\textit{right panel}) for hard-sphere model (\textit{solid curve}),
  accretion model for $\alpha=0$ (\textit{dashed curve}), and accretion
  model for $\alpha=2/3$ (\textit{short dashed curve}).   } 
  \label{fig:q_value_accretion.eps}
\end{figure}

We compare the accretion model with the hard sphere model in the laminar
 gas flow.
We performed simulations with the same model parameters except for
 collision models (models 100H0, 100AC0, and 100AS0) in order to check the
 validity of the accretion model. 
The initial stopping time of particles is $\tilde t_\mathrm{stop,0} = 1.0$.
In the accretion models, we calculate the drag coefficient using the
 constant model ($\alpha=0$) and the Stokes' law model ($\alpha=2/3$). 
The left panel of Figure \ref{fig:q_value_accretion.eps} shows the time
 evolution of $Q$.
In the phase where $Q$ decreases, there are no clear differences among
 all models. 
On the other hand, in the phase where $Q$ increases, the discrepancies appear among them.  
Toomre's $Q$ value in the accretion model with $\alpha=2/3$ is larger
 than those in the hard sphere model and the accretion model with
 $\alpha=0$.  
In the Stokes' law model, as particles grow, the drag coefficient
 becomes small and thus the dissipation becomes inefficient, which leads
 to the relatively large velocity dispersion.

The right panel of Figure \ref{fig:q_value_accretion.eps} shows the
 maximum mass of planetesimals for all models.
In the early stage, particles or aggregates do not grow.
There the difference in collision models is not important.
The difference can be seen in the late stage.
In the Stokes' law model $\alpha=2/3$, the drag coefficient changes
 as particles grow, which affects the evolution of the maximum mass.
In the hard-sphere model, the drag coefficient is constant although
 aggregates grow, therefore the evolution in the hard-sphere model must  
 be similar to that in the accretion model with $\alpha=0$.
However, the difference appears at $t/t_\mathrm{K} = 10$.
The maximum mass of the accretion model with $\alpha=0$ increases, but that of the hard-sphere model does not change.
At $t/t_\mathrm{K} = 10$, there are only a few particles or aggregates.
The maximum mass of planetesimals is sensitive to the initial condition and noise.
But, at $t/t_\mathrm{K} = 20$, the maximum mass converges to the same value.
The final mass of the planetesimal depends on the size of the computational domain \citep{Michikoshi2009}.
The dispersion of the mass of the final state is relatively small.

In the early stage where no planetesimals form, the time evolution is quite similar in all models. 
In the late stage, once planetesimals form, there are no remarkable differences among all models.
If we focus on the the evolution of statistical value, such as the velocity dispersion, the number of planetesimals,
we can investigate the gravitational instability and planetesimal formation using the accretion model.
Although the time evolution of the maximum mass of the planetesimal in the late stage depends on the initial noise, the maximum mass finally converge to the same value.
Therefore, we adopt the accretion model to investigate the gravitational instability from now on. 

\section{Results \label{sec:result}}
We performed 14 simulations with different disk models.  The model parameters are summarized in Table $\ref{table:model2}$.  
We fixed the following parameters: the optical depth, $\tau = 0.1$, the restitution coefficient $\epsilon = 0.01$, 
the ratio of the Hill radius to the diameter $\zeta=2.0$, the initial Toomre's $Q$ value $Q_{\mathrm{init}} = 3$, and the size of the computational domain $A=6$.
The basic dependence of planetesimal formation of these parameters was investigated in Papers I and II.

\begin{deluxetable}{ccccc}
  \tablecaption{Simulation parameters}
  \tablewidth{0pt}
  \tablehead{
  \colhead{Model} & \colhead{$\tilde t_{\mathrm{stop},0}$} & \colhead{Collision} & \colhead{$\alpha$} & \colhead{$\tilde v_\mathrm{dif}$}
  }
  \startdata 
010AS0 & 0.10     & Accretion      & 2/3   & 0   \\
025AS0 & 0.25     & Accretion      & 2/3   & 0   \\
050AS0 & 0.50     & Accretion      & 2/3   & 0   \\
100AS0 & 1.00     & Accretion      & 2/3   & 0   \\
200AS0 & 2.00     & Accretion      & 2/3   & 0   \\
400AS0 & 4.00     & Accretion      & 2/3   & 0   \\
1000AS0 & 10.00     & Accretion      & 2/3   & 0   \\
INFA   & $\infty$ & Accretion      & -     & 0   \\
100AC0 & 1.00     & Accretion      & 0   & 0   \\
100H0 & 1.00     & Hard & 0   & 0   \\
100AC1 & 1.00     & Accretion & 0   & 10 \\
100AS1 & 1.00     & Accretion & 2/3 & 10 \\
100AC2 & 1.00     & Accretion & 0   & 20 \\
100AS2 & 1.00     & Accretion & 2/3 & 20 \\
  \enddata
  \tablecomments{ Parameters $\tilde t_{\mathrm{stop},0}$, $\alpha$, and $v_{\mathrm{dif}}$  are the stopping time of particles,
    the power-law index for the stopping time $\tilde t_{\mathrm{stop},i} \propto \tilde m_i^{\alpha}$, and the velocity difference from Kepler velocity.  
    We fixed other parameters: the optical depth $\tau = 0.1$, the restitution coefficient $\epsilon = 0.01$, the ratio of the Hill radius to the diameter $\zeta=2.0$, the initial $Q$ value $Q_\mathrm{init}=3$, and the size of the computational domain $A=6$.}
\label{table:model2}
\end{deluxetable}

\subsection{Time evolution}
\begin{figure}
	\begin{center}
		\begin{tabular}{cccc}
			 &	
			\resizebox{37mm}{!}{\includegraphics{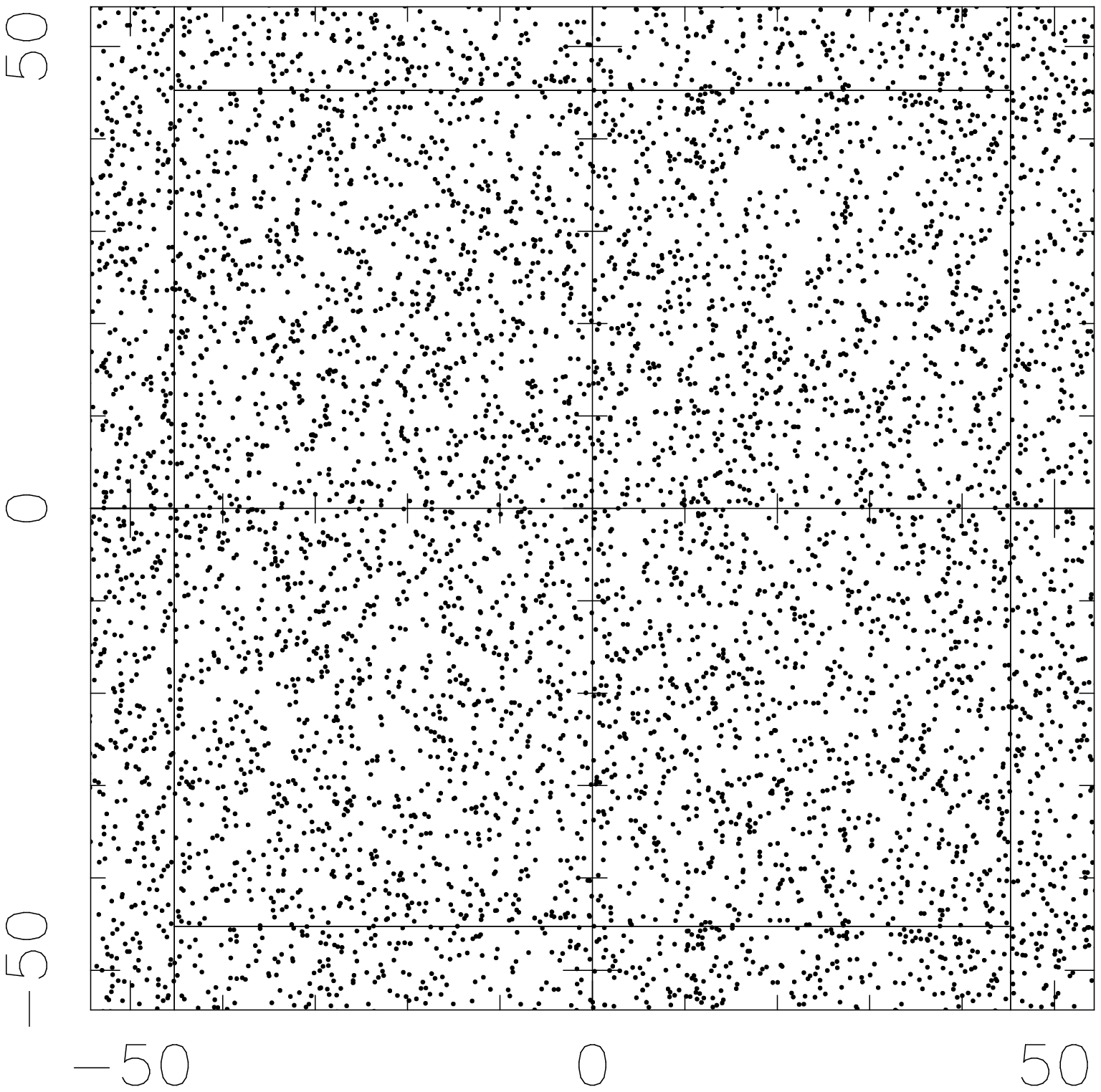}} &
			\resizebox{37mm}{!}{\includegraphics{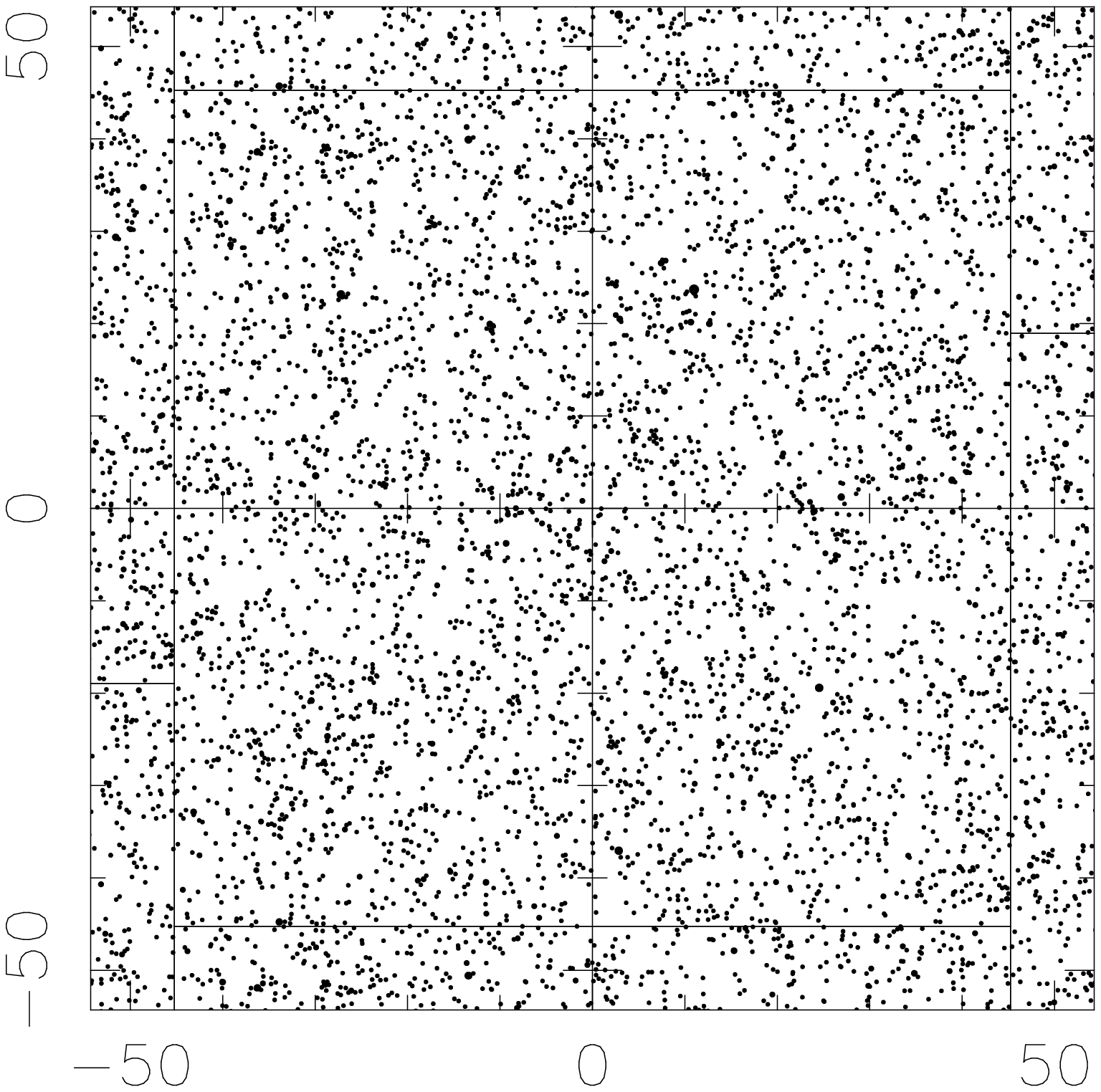}} &
			\resizebox{37mm}{!}{\includegraphics{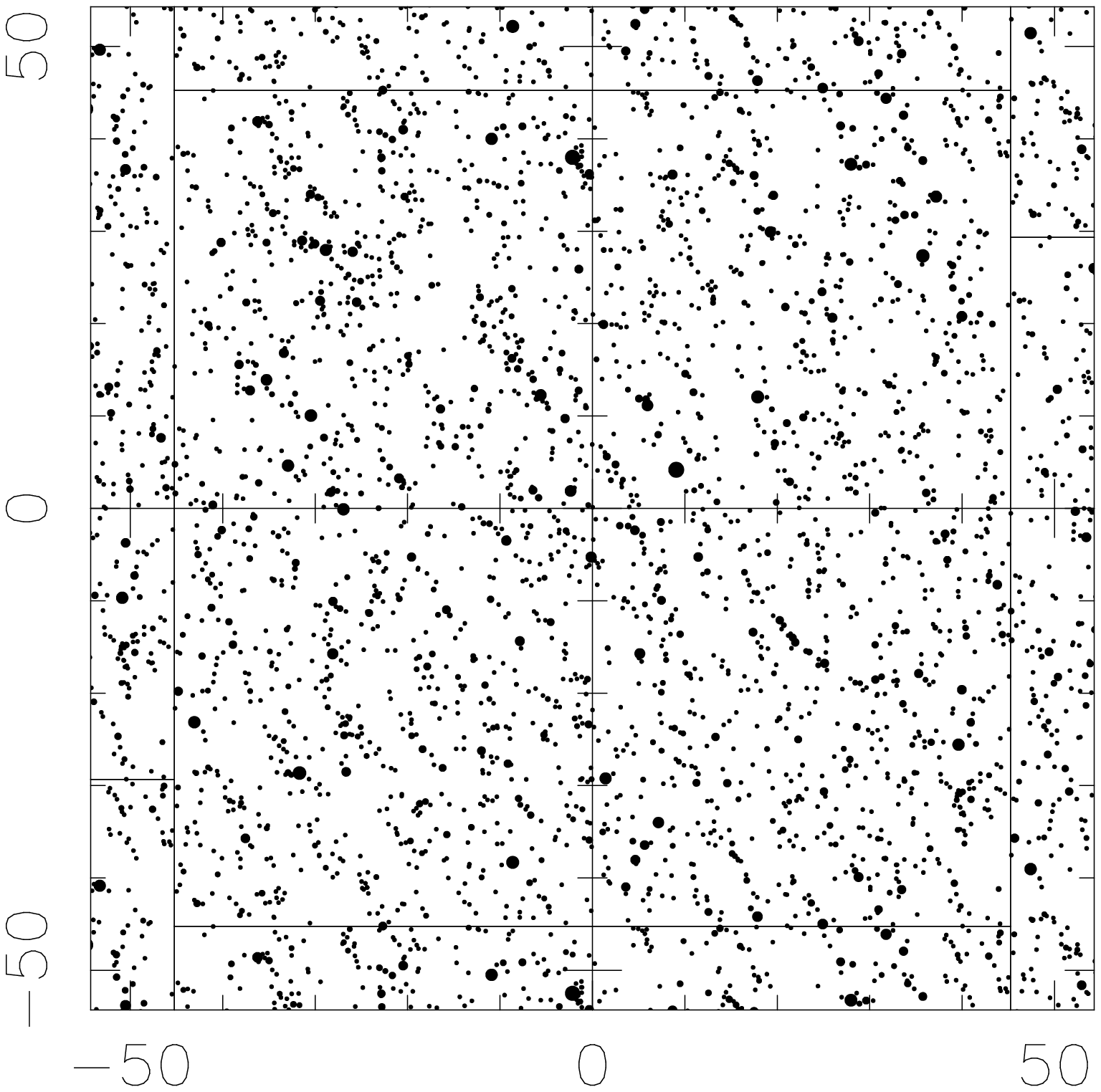}} \\
			 \multirow{1}{*}{$\tilde y$} 
			&
			\resizebox{37mm}{!}{\includegraphics{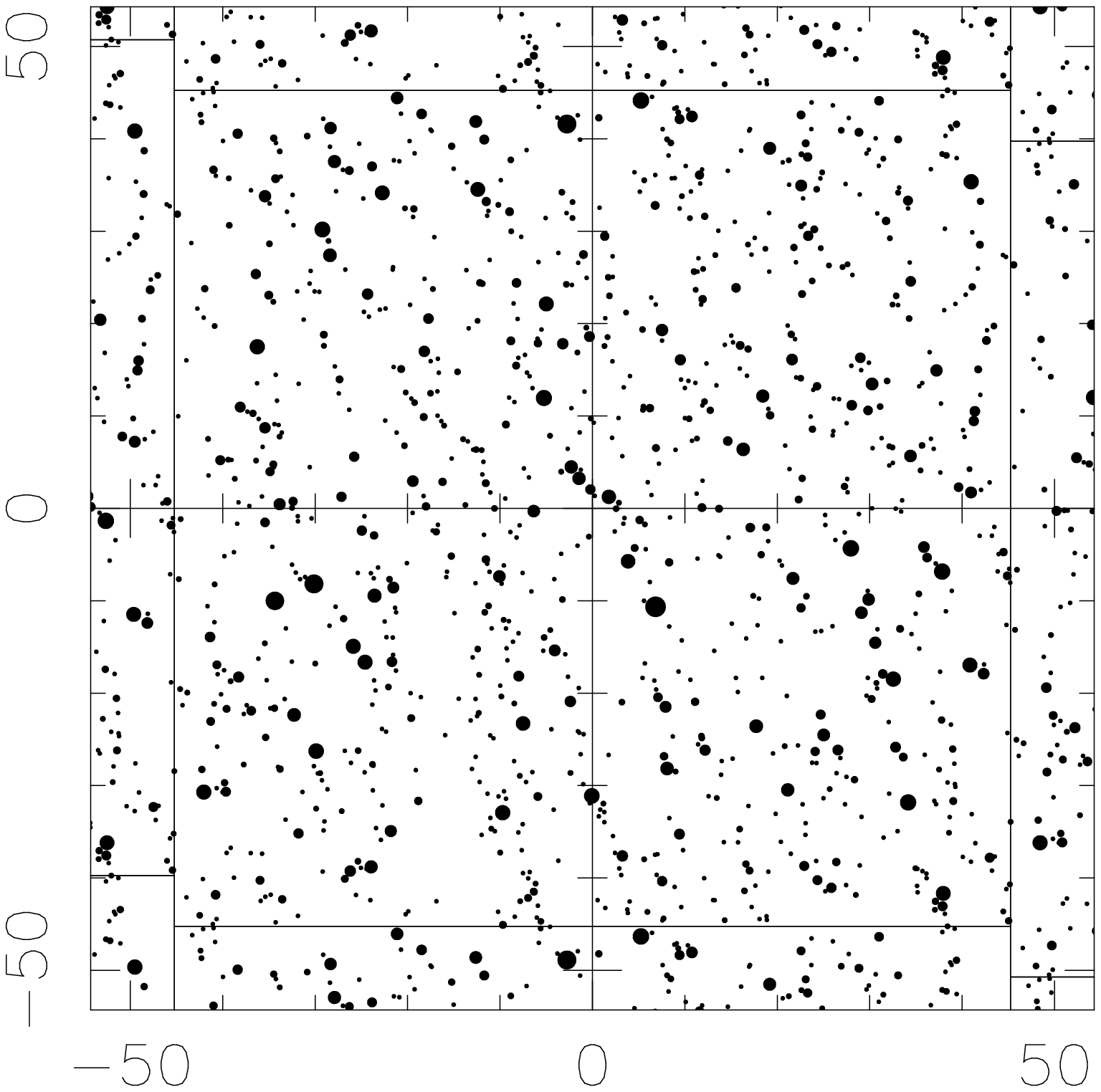}} &
			\resizebox{37mm}{!}{\includegraphics{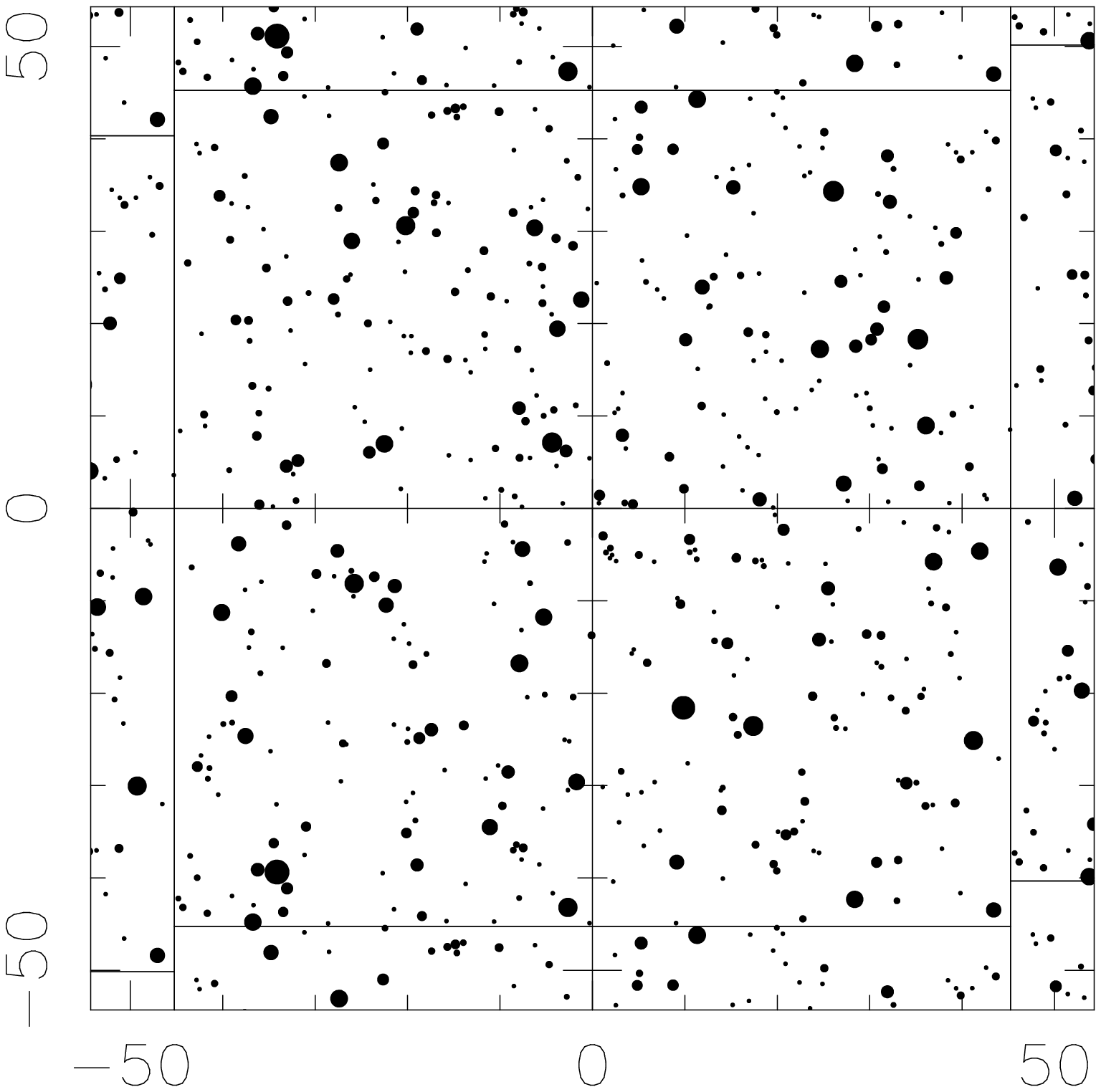}} &
			\resizebox{37mm}{!}{\includegraphics{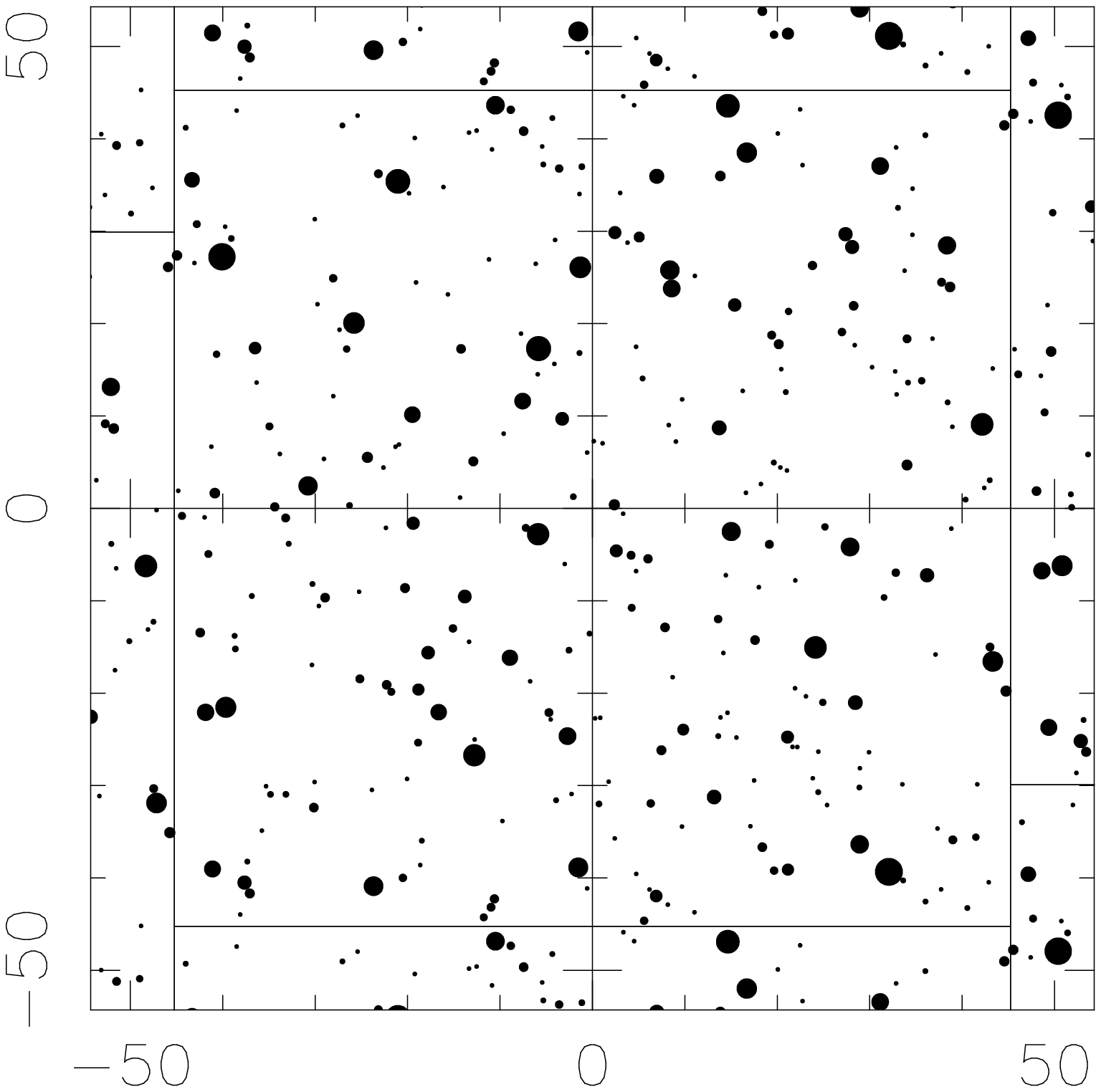}} \\      
			&
			\resizebox{37mm}{!}{\includegraphics{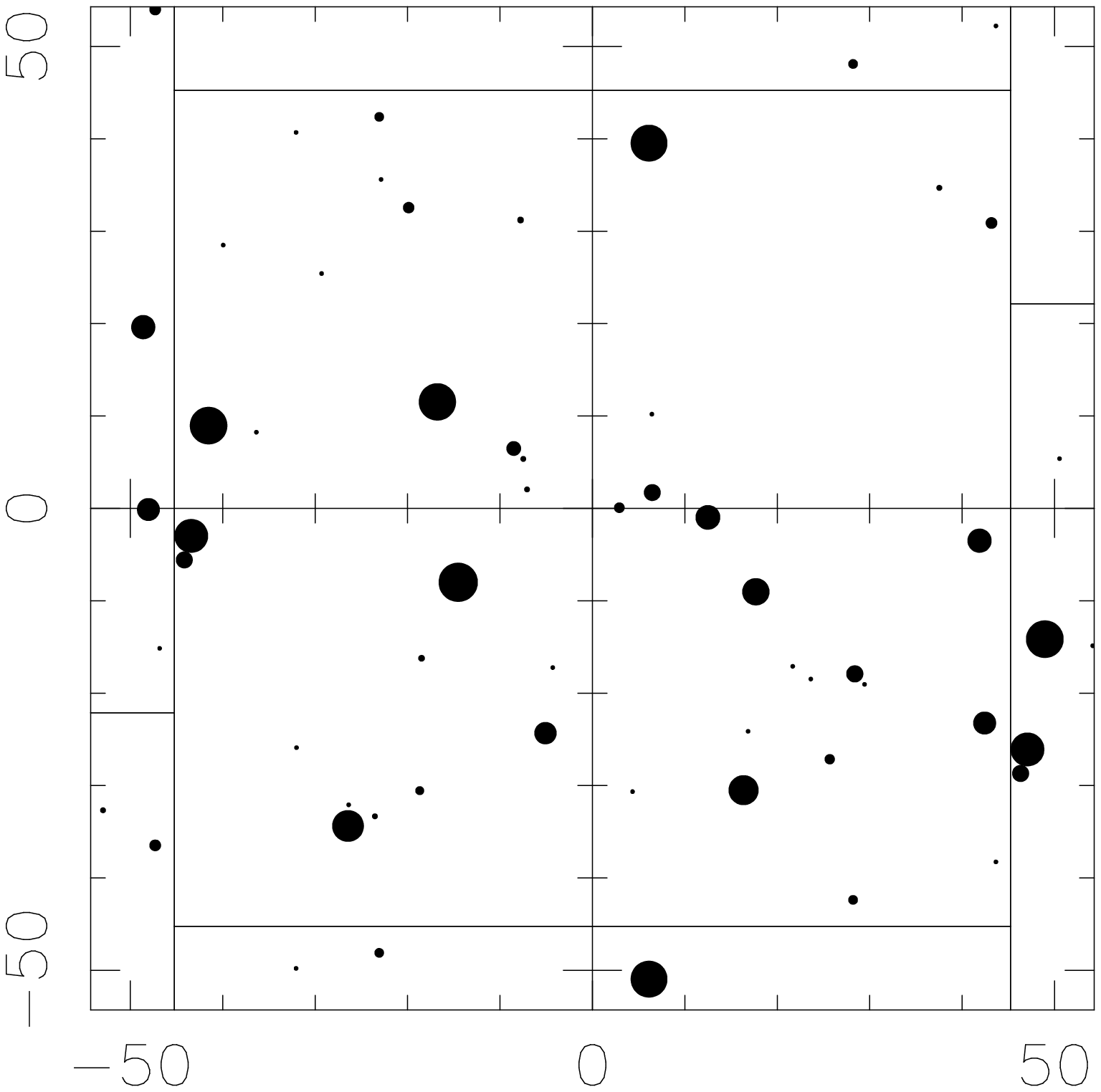}} &
			\resizebox{37mm}{!}{\includegraphics{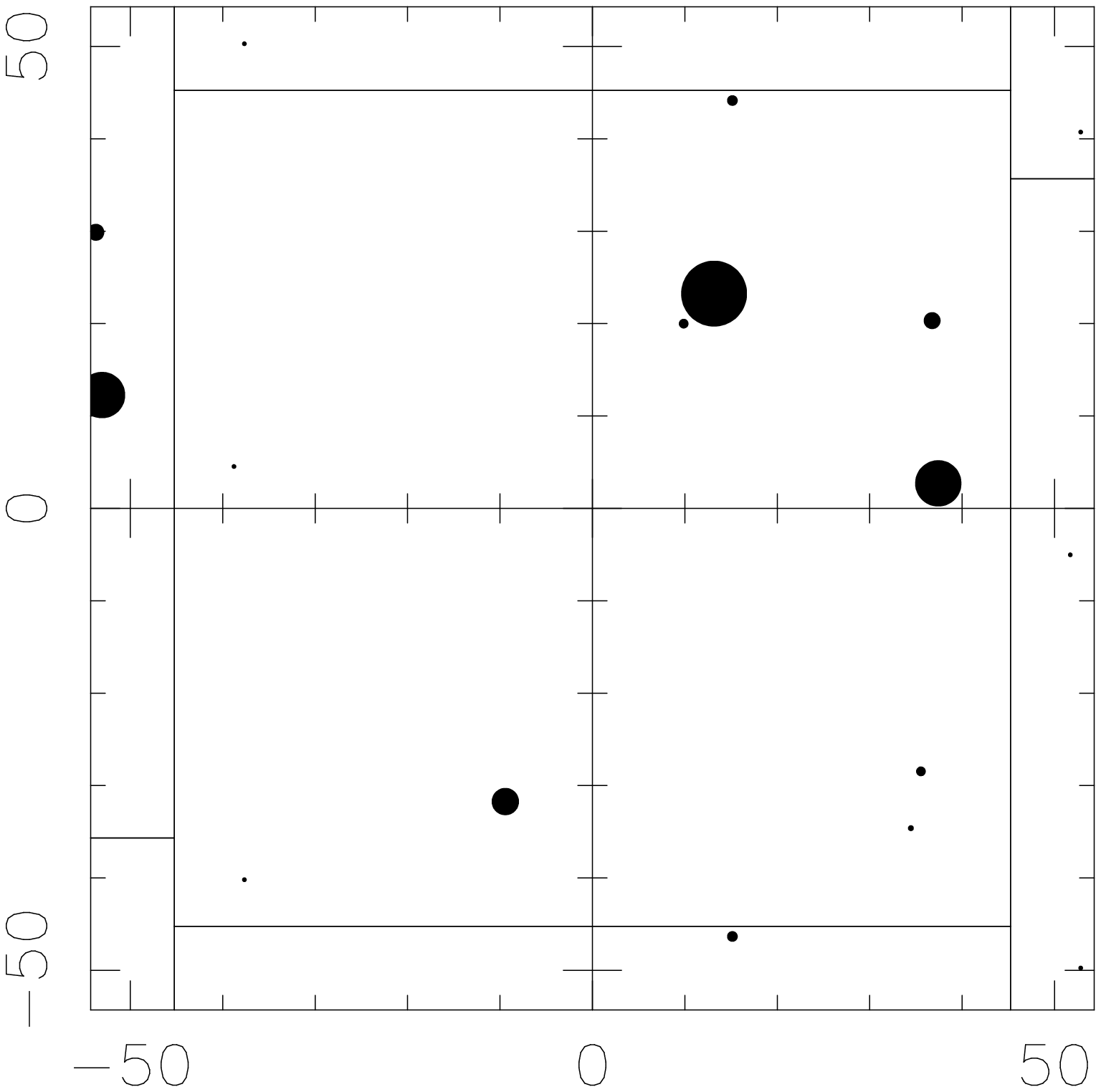}} &
			\resizebox{37mm}{!}{\includegraphics{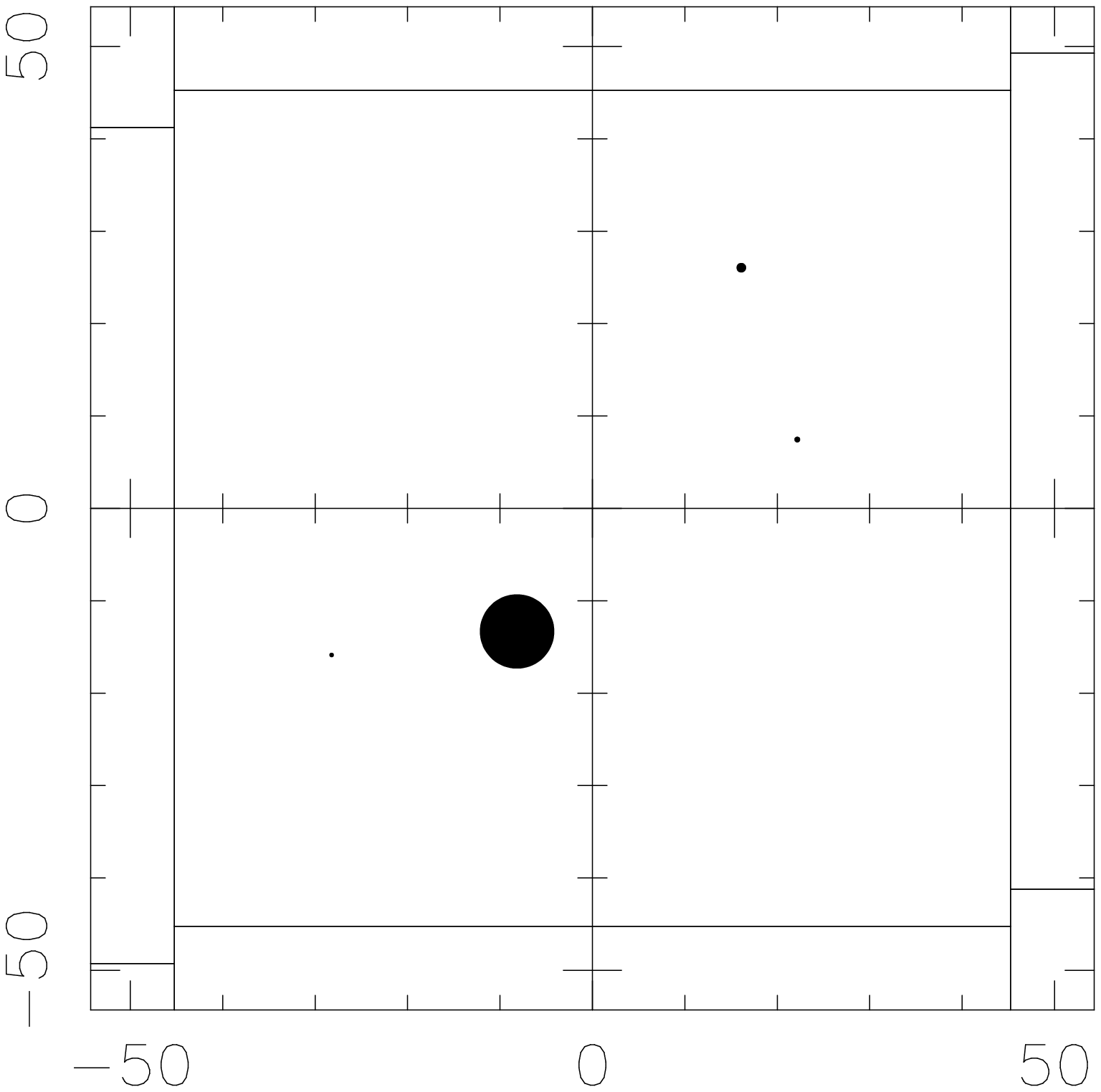}} \\
			& & $\tilde x$ & \\
		\end{tabular}
		\caption{Spatial distribution of particles in the $\tilde x \tilde y$ plane in in the model 100AS0 at $t/t_\mathrm{K}=0.0, 0.2, 0.4, 0.6, 0.8, 1.0, 2.0, 4.0,$ and $6.0$
			from left to right and from top to bottom.
          Circles represent particles and their size is proportional to the physical size of particles.  }
		\label{fig:snap_shot}
	\end{center}
\end{figure}

Figure \ref{fig:snap_shot} shows the typical evolution of the simulation (model 100AS0).
At $t/t_\mathrm{K}=0$, and $0.2$, particles are distributed randomly and uniformly in the computational box.
No gravitational instability occurs, thus we can see no remarkable structures. 
At $t/t_\mathrm{K}=0.4$, the large non-axisymmetric density structures appear.
Then, the gravitational instability seems to start.
In the model where ($t_\mathrm{stop}=0.25$), the structure is not clear.
For the short stopping time models, the velocity dispersion is small and the critical wave length is short.
Thus, we cannot observe the large density structure clearly. 
At $t/t_\mathrm{K}=0.6$, particles begin to grow in the dense region. 
At $t/t_\mathrm{K}=0.8,$ and $1.0$, particles continue to grow, and the number of small particles decreases rapidly.
The small particles are absorbed by large particles.
The rapid collisional growth of particles continues until the one large planetesimal finally forms.

The formation process of planetesimals in the laminar gas model is similar to those in the gas-free model (papers I and II).
The gas drag from the laminar gas  does not qualitatively change formation process.
This process is divided into three stages: the formation of density structures, the creation of planetesimal seeds, and their collisional growth. 

The linear stability analysis of gas-free gravitational instability shows that the disk is stable when $Q>1$ \citep{Toomre1964}.
As discussed in \S 3, the dust layer in the laminar gas disk is secularly unstable although $Q>1$. 
However, the growth time of the dissipative gravitational instability is longer than the sedimentation in the initial stage.  
Thus, the disk shrinks vertically.
As the velocity dispersion and the scale height decrease, the growth rate of the dissipative gravitational instability increases.
Then, the gravitational instability becomes a dominant process, and the wake-like structures appear.  
These structures fragment into the seeds of planetesimals. 
The seeds grow rapidly owing to mutual collisions.
Finally, almost all mass in the computational domain is absorbed by only one planetesimal in this calculation.
The size of the final planetesimal depends on the size of the computational domain (paper II).

\subsection{Effect of Stopping Time}
We assume that the difference in the rotational velocity from the Kepler velocity is equal to zero ($v_\mathrm{dif}=0$)  (models INFA, 010AS0 - 1000AS0).
In the model INFA, $t_{\mathrm{stop},0}=\infty$, i.e., the drag term is neglected.
In other models, the stopping times are $\tilde t_{\mathrm{stop},0} = 0.1, 0.25, 0.5, 1.0, 2.0, 4.0,$ and $10.0$.
We adopt the Stokes' law model, $\alpha = 2/3$.

\subsubsection{Toomre's $Q$ value \label{sec:qvalue}}
\begin{figure}
  \begin{center}
	\plotone{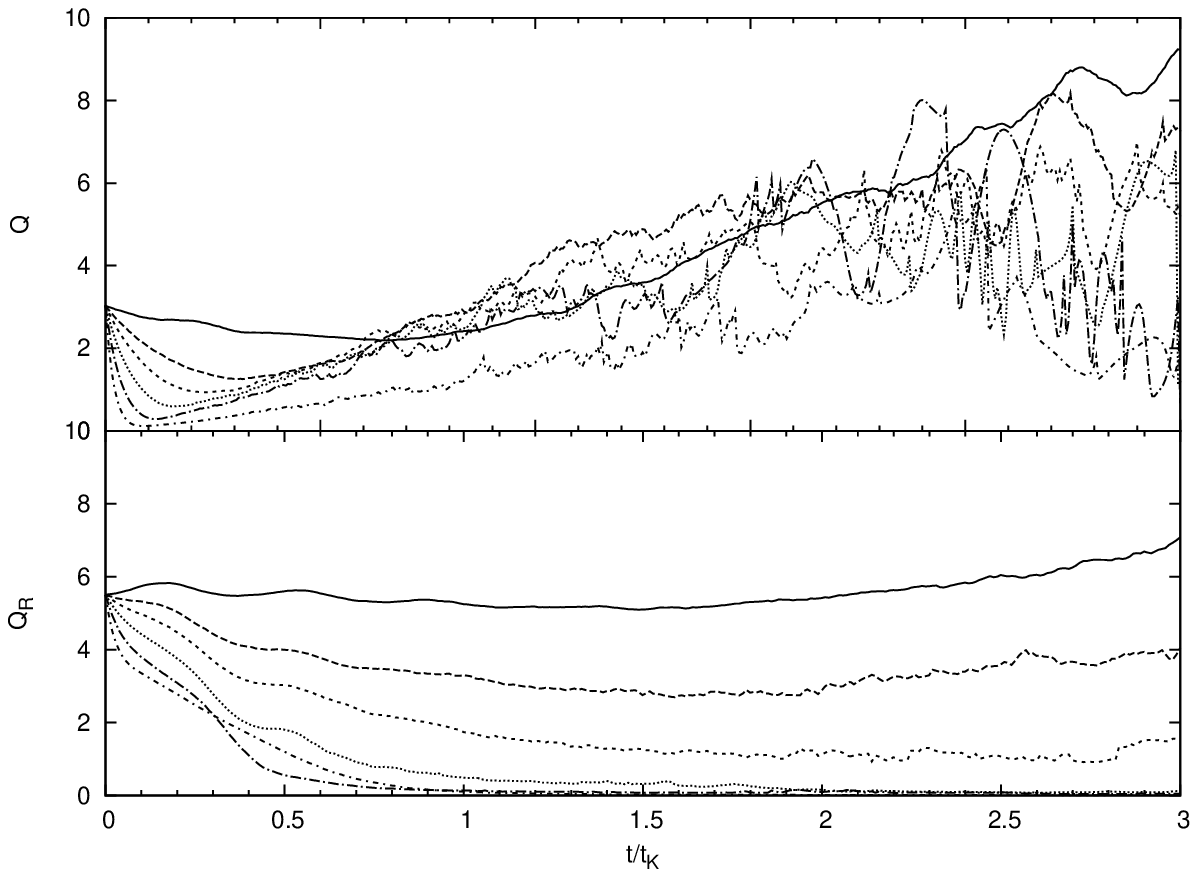}
  \end{center}
\caption{The time evolutions of Toomre's $Q$ value (\textit {top panel}) and the normalized scale height (\textit{bottom panel}) for gas-free model (\textit{solid curve}), $\tilde t_\mathrm{stop,0} = 4.0$ (\textit{dashed curve}), $\tilde t_\mathrm{stop,0} = 2.0$ (\textit{short dashed curve}) , $\tilde t_\mathrm{stop,0} = 1.0$ (\textit{dotted curve}) , $\tilde t_\mathrm{stop,0} = 1.0$ (\textit{dash-dotted curve}), and $\tilde t_\mathrm{stop,0} = 0.5$ (\textit{dot-short-dashed curve}) (models INFA, 400AS0, 200AS0, 100AS0, 050AS0, and 025AS0).  }
  \label{fig:q_value_hard.eps}
\end{figure}

The top panel of Figure \ref{fig:q_value_hard.eps} shows the time evolution of Toomre's $Q$ values.  
In all models, Toomre's $Q$ value initially decreases.  
In the gas-free model (model INFA), the kinetic energy is dissipated only by inelastic collisions. 
The dissipation timescale due to inelastic collisions is approximately estimated by $t_\mathrm{c} \simeq 1/(\tau \Omega (1-\epsilon^2)) \simeq 10.0\Omega^{-1}$ \citep{Goldreich1978}.  
In the laminar gas models, the decrease of
Toomre's $Q$ value is caused by the dissipation due to both the gas drag and inelastic collisions.  

\begin{figure}
  \begin{center}
	\plotone{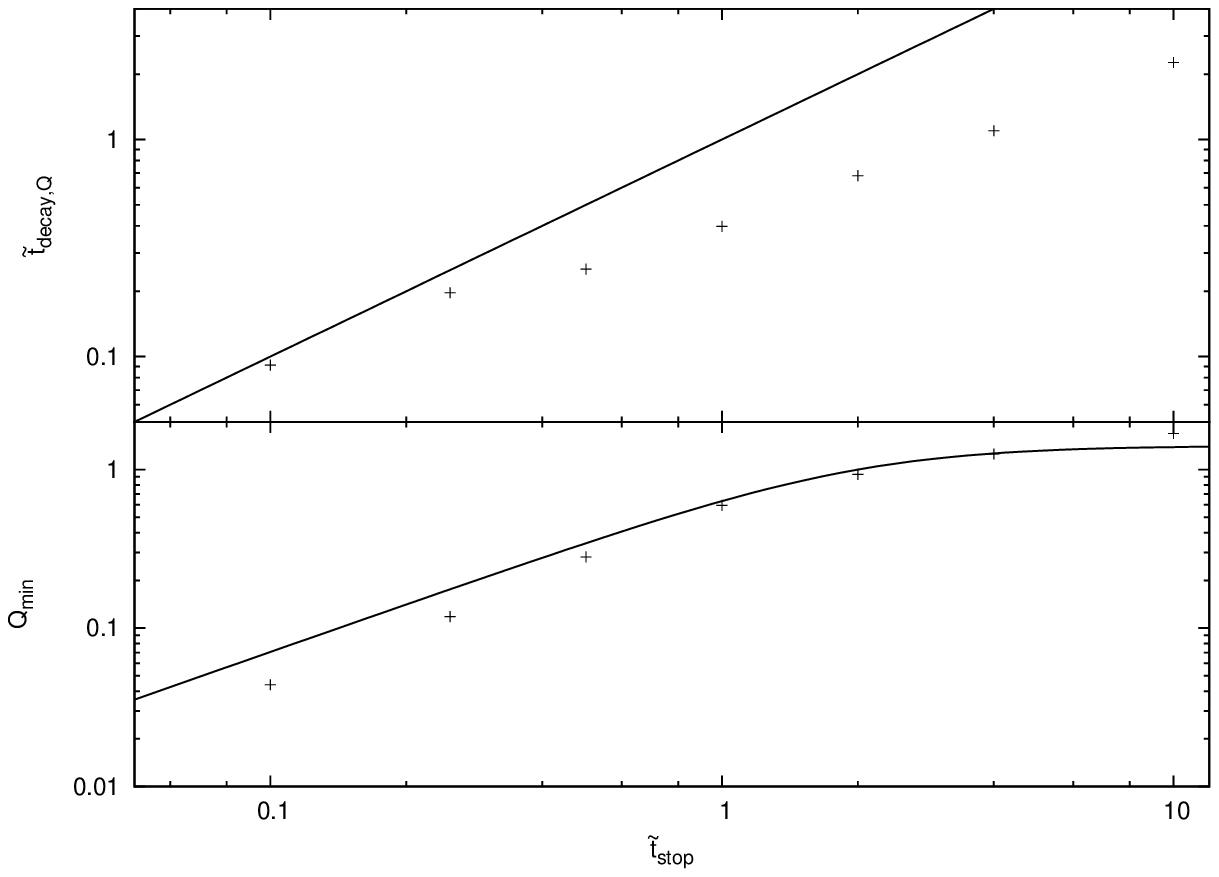}
  \end{center}
  \caption{The decay time $\tilde t_\mathrm{decay,Q}$ of the Toomre's $Q$ value (\textit{top panel}) and the minimum $Q$ value $Q_\mathrm{min}$ (\textit{bottom panel}) as a function of $\tilde t_\mathrm{stop,0}$.
  The cross denotes the result of Numerical simulations (models 010AS0, 025AS0, 050AS0, 100AS0, 200AS0, 400AS0, and 1000AS0). 
  In the top panel, the solid line shows the line $\tilde t_\mathrm{decay,Q} = \tilde t_\mathrm{stop,0}$.
  In the bottom panel, the solid line shows the analytical estimation of the minimum Toomre's $Q$ value expressed by Equation (\ref{eq:qmin}). 
 }
  \label{fig:min_q_value.eps}
\end{figure}

The top panel of Figure \ref{fig:min_q_value.eps} shows the dependence of the decay time $t_\mathrm{decay,Q}$ on $t_\mathrm{stop,0}$, where 
the decay time of Toomre's $Q$ value $t_\mathrm{decay,Q}$ is defined by
\begin{equation}
  t_\mathrm{decay,Q} = t_\mathrm{min,Q}\frac{Q_\mathrm{init}}{Q_\mathrm{init}-Q_\mathrm{min}},
\end{equation}
where $Q_\mathrm{min}$ is the minimum $Q$ value at $t=t_\mathrm{min,Q}$.
The decay time $t_\mathrm{decay,Q}$ is proportional to $t_\mathrm{stop,0}$ for $\tilde t_\mathrm{stop,0} < 1$.
Toomre's $Q$ value is proportional to the velocity dispersion $\sigma_x$.
The inelastic collisions and the gas drag decrease the velocity dispersion.
If the drag is sufficiently strong, the decay time scale of the velocity dispersion is on the order of the stopping time.
The timescale of the decay of the velocity dispersion due to inelastic collisions is about $t_\mathrm{c} \simeq  10.0 \Omega^{-1}$.
Thus, when $\tilde t_\mathrm{stop,0}>1$  as $\tilde t_\mathrm{stop,0}$ increases, we cannot neglect collisions.
Therefore the decay timescale of the velocity dispersion is smaller than the stopping time for $\tilde t_\mathrm{stop,0}>1$.

In all models, Toomre's $Q$ value has the minimum value $Q_\mathrm{min}$ at $t_\mathrm{min,Q}$.  
In the gas-free model, the minimum $Q$ value is about $2$ (papers I, and II), and according to the linear theory,
the critical value of a thin disk is $Q=1$.
This is because the decrease in the velocity dispersion due to inelastic collisions is sufficiently slow.
However, as shown in Figure \ref{fig:q_value_hard.eps}, if the gas drag is strong, $Q$ value becomes smaller than unity.
As the stopping time shortens, the minimum $Q$ value decreases. 
For example, the minimum $Q$ value is 0.15 for $\tilde t_\mathrm{stop,0} = 0.25$.

The axisymmetric structure starts forming when $Q$ value is minimum.
For example, in the model where $\tilde t_\mathrm{stop,0}=1.0$, $Q$ value is minimum at $t/t_\mathrm{K}=0.3$, and we can see the 
non-axisymmetric structure at $t/t_\mathrm{K}=0.35$.
Particles are scattered by the non-axisymmetric structures, and the velocity dispersion of particles increases.

If the decrease of the $Q$ value is more rapid than the gravitational instability, the dust
layer cannot collapse although the dust layer is gravitationally unstable.
As $Q$ value decreases, the timescale of the gravitational instability becomes short, 
therefore, the critical $Q$ value is determined by $t_\mathrm{stop,0} = t_\mathrm{GI}$, where $t_\mathrm{GI}$ is the timescale of the gravitational instability.
Here, we assume $t_\mathrm{stop,i} = t_\mathrm{stop,0}$ because particles do not grow much before gravitational instability.
From Equation (\ref{eq:crit_Q}), we obtain the following condition:
\begin{equation}
  Q_\mathrm{min}= \sqrt{\frac{2\tilde t_\mathrm{stop,0}^2}{4 +\tilde t_\mathrm{stop,0}^2}}.
\label{eq:qmin}
\end{equation}
The minimum Toomre's $Q$ value as a function of $t_\mathrm{stop,0}$ is shown in the bottom panel of Figure \ref{fig:min_q_value.eps}.
Strictly speaking, Equation (\ref{eq:qmin}) is valid only when the disk is thin.
Nevertheless, the analytical expression agrees with the numerical result.

\subsubsection{\label{sec:scaleheight}Roche Density}
We define the scale height of a dust layer as the mean square root of $z$, $h = \sqrt{2} \langle z_i^2 \rangle^{1/2}$.  We can calculate the mean density of the dust layer from the scale height, $\rho=\Sigma/h$.
The density at which the self-gravity is equal to the tidal force is called the Roche density, which is defined by the following equation \citep[e.g.,][]{Yamoto2004}:
\begin{equation}
  \rho_\mathrm{R} = \frac{9}{4 \pi} \frac{M_\mathrm{s}}{a_0^3}.
\end{equation}
The dust layer may collapse if the dust layer density exceeds the Roche density.
The linear analysis gives the more precise criterion \citep{Sekiya1983, Yamoto2004}.
However, the criterion does not change very much.
Thus, we introduce $Q_\mathrm{R}$ given by 
\begin{equation}
  Q_\mathrm{R} = \frac{\rho_\mathrm{R}}{\rho} = \frac{9}{4\pi} \frac{h \Omega^2}{G \Sigma },
\end{equation}
which corresponds to the non-dimensional scale height.

\begin{figure}
  \begin{center}
	\plotone{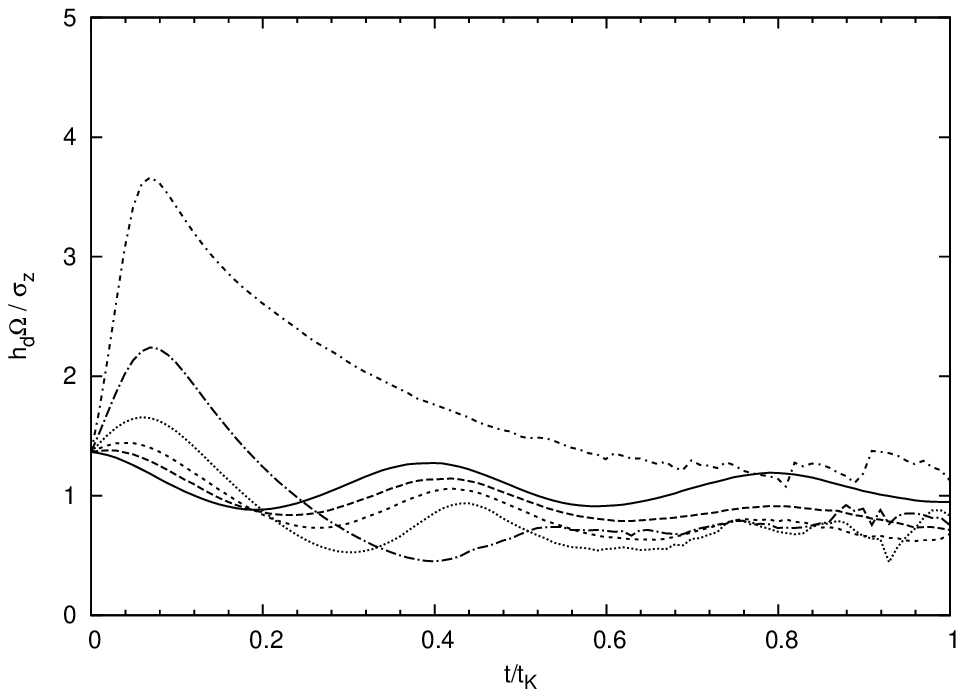}
  \end{center} 
\caption{The time evolutions of $\tilde h / \tilde \sigma$ for gas-free model (\textit{solid curve}), $\tilde t_\mathrm{stop,0} = 4.0$ (\textit{dashed curve}), $\tilde t_\mathrm{stop,0} = 2.0$ (\textit{short dashed curve}) , $\tilde t_\mathrm{stop,0} = 1.0$ (\textit{dotted curve}) , $\tilde t_\mathrm{stop,0} = 0.5$ (\textit{dash-dotted curve}), and $\tilde t_\mathrm{stop,0} = 0.25$ (\textit{dot-short-dashed curve}) (models INFA, 400AS0, 200AS0, 100AS0, 050AS0, and 025AS0). }
  \label{fig:v_over_h}
\end{figure}

The bottom panel of Figure \ref{fig:q_value_hard.eps} shows the time evolution of $Q_R$. 
The time evolution of $Q_R$ is similar to Toomre's $Q$.
The value $Q_R$ has the minimum value $h_\mathrm{min}$ at $t_\mathrm{min,h}$ in all models.  
But, the time evolution of $Q_R$ is slower than that of $Q$ value.
This result indicates that the standard hydrostatic relation $h \simeq \sqrt{2} \sigma_z / \Omega$ is not satisfied.
Figure \ref{fig:v_over_h} shows the time evolution of the ratio $h / \sigma_z$, which is on the order of the sound crossing time across a scale height.
In the gas-free model, the ratio $h /\sigma_z$ is nearly constant in $t/t_\mathrm{K}<1$.
The hydrostatic relation is satisfied.
In a laminar gas model , the ratio $h / \sigma_z$ is not constant.
As the stopping time decreases, the variation of the ratio becomes steep.
Especially, those for $t<1 t_\mathrm{K}$ is prominent.
In the model $t_\mathrm{stop}=0.25$, the ratio $h / \sigma_z$ increases for $t < 0.08 t_\mathrm{K}$.
The maximum value is about $3.7$.
The sound crossing time is $h / \sigma_z/\simeq 1.4$ in the gas-free model.
When $\tilde t_\mathrm{stop,0} < 1$, the dissipation is faster than the sound crossing.
Therefore, the velocity dispersion decays before the scale height decays, and the ratio becomes larger than unity.
In Figure \ref{fig:v_over_h}, we can see the oscillation. 
The period is about $0.4 t_\mathrm{K}$, and the amplitude decreases gradually.
The reason for the oscillation is not clear.  This may be due to some kind of the relaxation process.

We cannot use the Roche density as the collapse criterion for $\tilde t_\mathrm{stop}>1$.
For $\tilde t_\mathrm{stop,0}\ge 4.0$, $Q_R$ is always larger than unity. 
But the large particles form.
For $\tilde t_\mathrm{stop,0}=2.0$, $Q_R$ becomes unity at $t/t_\mathrm{K}=2$, but the large particles start forming at $t/t_\mathrm{K}=0.6$.
For $\tilde t_\mathrm{stop}\le 1$, the time when $Q_R=1$ corresponds to the time when the large particles start forming.
When we derive the Roche criterion, we simply compare the self-gravity with the tidal force.  
However, we showed that the non-axisymmetric structure develops.  
The azimuthal motion may be important to discuss the formation of the wake-like structure.
Therefore, the Roche criterion is not applicable.

\begin{figure}
  \begin{center}
	\plotone{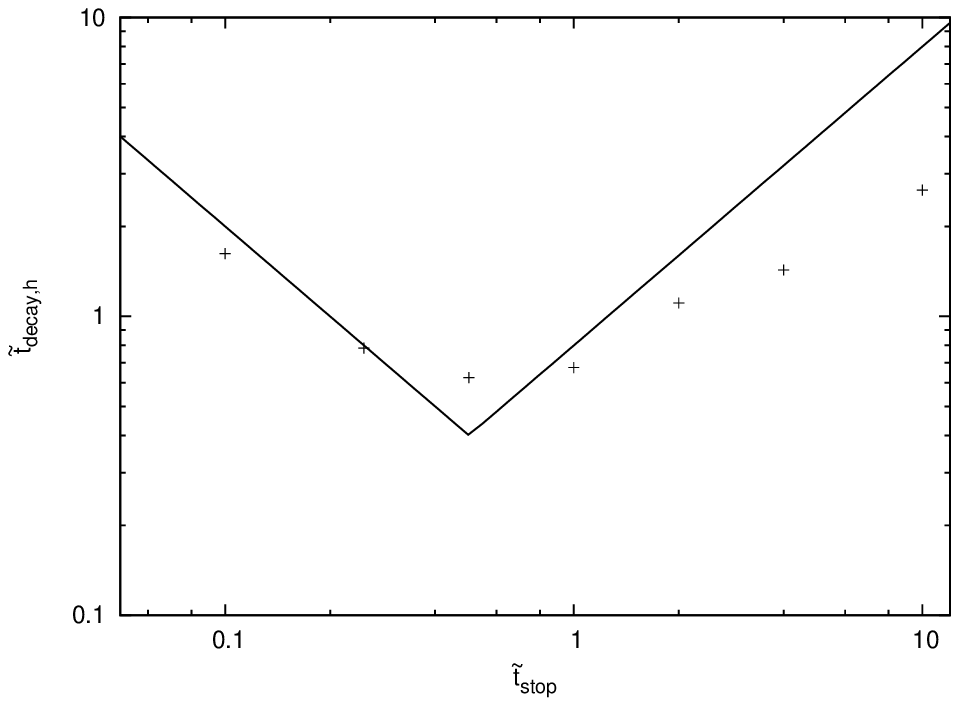}
  \end{center}
\caption{ The decay time of the scale height as a function of $\tilde t_\mathrm{stop,0}$.
  The cross denotes the result of Numerical simulations (models 010AS0, 025AS0, 050AS0, 100AS0, 200AS0, 400AS0, and 1000AS0).  }
  \label{fig:min_height_time.eps}
\end{figure}

The decay time of $Q_R$ value  $t_\mathrm{decay,h}$ is defined by
\begin{equation}
  t_\mathrm{decay,h} = t_\mathrm{min,h}\frac{h_\mathrm{init}}{h_\mathrm{init}-h_\mathrm{min}},
\end{equation}
where $h_\mathrm{init}$ is the initial scale height of a dust layer.
Figure \ref{fig:min_height_time.eps} shows $t_\mathrm{decay,h}$ as a function of $t_\mathrm{stop,0}$.
The decay time of $Q_R$ has a minimum value at $\tilde t_\mathrm{stop,0} = 0.5$.
For $\tilde t_\mathrm{stop,0} < 0.5$, the decay time shortens with the stopping time.
For $\tilde t_\mathrm{stop,0} < 0.5$, the strong coupled particles with gas fall to the midplane at the terminal velocity.
The settling time is proportional to $\tilde t_\mathrm{stop,0}^{-1}$.
On the other hand, for $\tilde t_\mathrm{stop,0} > 0.5$, the decay time lengthens with longer stopping time.
In this regime, the particles oscillate around the equatorial plane \citep{Nakagawa1986}.
The decay time scale of the amplitude is proportional to the stopping time.

\subsubsection{Number of Planetesimal Seeds}

\begin{figure}
  \begin{center}
	\plotone{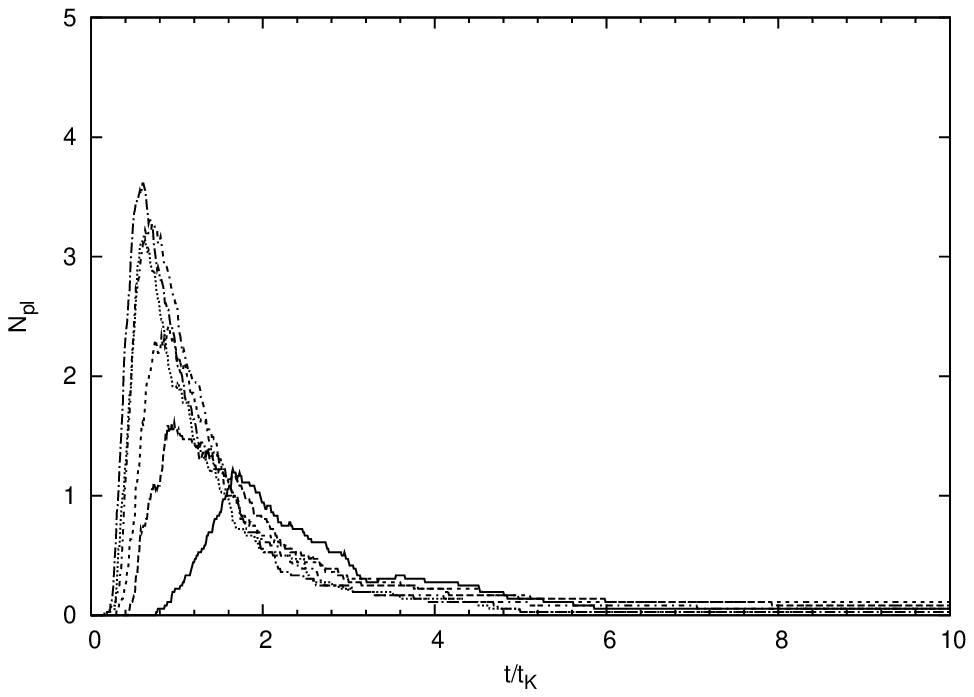}
  \end{center}
  \caption{
The time evolutions of the number of the planetesimals for gas-free model (\textit{solid curve}), $\tilde t_\mathrm{stop,0} = 4.0$ (\textit{dashed curve}), $\tilde t_\mathrm{stop,0} = 2.0$ (\textit{short dashed curve}) , $\tilde t_\mathrm{stop,0} = 1.0$ (\textit{dotted curve}) , $\tilde t_\mathrm{stop,0} = 0.5$ (\textit{dash-dotted curve}), and $\tilde t_\mathrm{stop,0} = 0.25$ (\textit{dot-short-dashed curve}) (models INFA, 400AS0, 200AS0, 100AS0, 050AS0, and 025AS0).  }
  \label{fig:cluster_number.eps}
\end{figure}

Figure \ref{fig:cluster_number.eps} shows the time evolution of the number of planetesimal seeds per the area $\lambda_\mathrm{m}^2$, $N_\mathrm{pl}$.  
We define a particle whose mass is larger than $M_\mathrm{linear}$ as a seed of planetesimals, where $M_\mathrm{linear}$ is the planetesimal mass predicted by the linear theory $\pi \Sigma (\lambda_\mathrm{m}/2)^2$.
In all models, $N_\mathrm{pl}$ has a maximum value at $t/t_\mathrm{K} = 0.5-1$. The number $N_\mathrm{pl}$ increases rapidly in the early stage, and decreases in the late stage.
In the early stage, gravitational instability occurs and the small density fluctuation grows and many planetesimal seeds form.
In this stage, the number of seeds increases.
They form through the gravitational instability.
Once many planetesimal seeds form, the mutual collisions among them start. The number of planetesimals decreases rapidly in the late stage.
In the final state, almost all particles are absorbed by a few particles.
The number of planetesimals in the final state depends on the size of the computational domain (paper II).

\begin{figure}
  \begin{center}
	\plotone{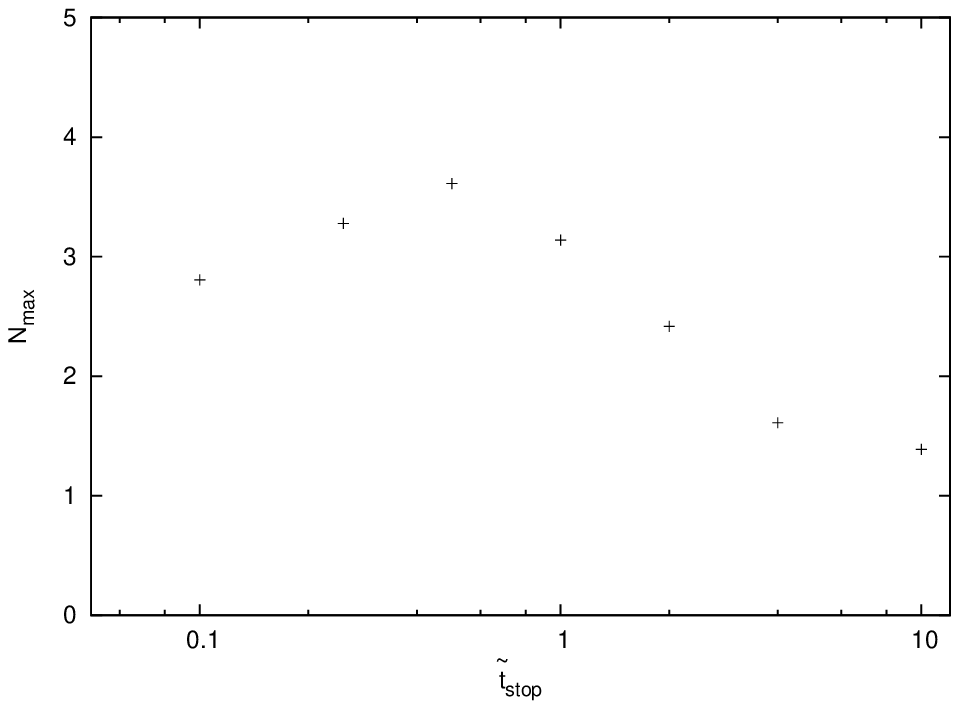}
  \end{center}
\caption{ The maximum number of the planetesimals as a function of $\tilde t_\mathrm{stop,0}$.
  The cross denotes the result of Numerical simulations (models 010AS0, 025AS0, 050AS0, 100AS0, 200AS0, 400AS0, and 1000AS0).  }
  \label{fig:max_number.eps}
\end{figure}

The peak of $N_\mathrm{pl}$ depends on the stopping time $\tilde t_\mathrm{stop,0}$.
Figure \ref{fig:max_number.eps} shows the maximum number of the planetesimal $N_\mathrm{max}$ as a function of the stopping time $\tilde t_\mathrm{stop,0}$.
The maximum number of planetesimals $N_\mathrm{max}$ has a peak at $\tilde t_\mathrm{stop,0} \simeq 0.5$.
As discussed in \S \ref{sec:scaleheight}, the scale height has the minimum value at $\tilde t_\mathrm{stop,0} \simeq 0.5$.
The characteristic length scale of gravitational instability corresponds to the scale height.
As the characteristic length of gravitational instability shortens, the number of the planetesimals formed increases.

\subsubsection{Mass of Planetesimals}

\begin{figure}
  \begin{center}
	\plotone{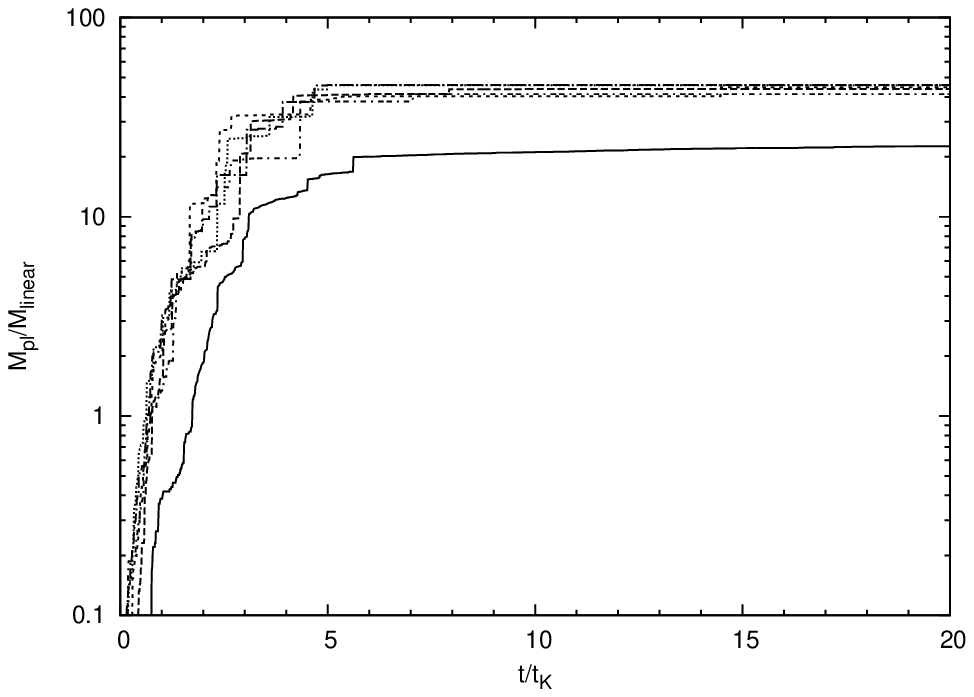}
  \end{center}

\caption{The time evolutions of the mass of the largest planetesimal for gas-free model (\textit{solid curve}), $\tilde t_\mathrm{stop,0} = 4.0$ (\textit{dashed curve}), $\tilde t_\mathrm{stop,0} = 2.0$ (\textit{short dashed curve}) , $\tilde t_\mathrm{stop,0} = 1.0$ (\textit{dotted curve}) , $\tilde t_\mathrm{stop,0} = 0.5$ (\textit{dash-dotted curve}), and $\tilde t_\mathrm{stop,0} = 0.25$ (\textit{dot-short-dashed curve}) (models INFA, 400AS0, 200AS0, 100AS0, 050AS0, and 025AS0). }
  \label{fig:largest_planetesimal.eps}
\end{figure}

Figure \ref{fig:largest_planetesimal.eps} shows the time evolution of the mass of the largest planetesimal.
In all models, the largest planetesimal grows rapidly in $t/t_\mathrm{K}<5$, and the growth is stalled at $t/t_\mathrm{K}\simeq 5-6$.
The growth in the gas-free model is slower than those in the gas models.
As discussed in \S \ref{sec:qvalue}, the gas drag enhances the dissipation of the kinetic energy of dust.
As the drag becomes stronger, the gravitational instability occurs earlier and the growth becomes more rapid.

The final mass of the largest planetesimal is $30-50 M_\mathrm{linear}$.
The clear dependence of the largest mass on the initial stopping time cannot be seen.
As seeds grow, the stopping time lengthens.
In the final state, we can neglect the effect of the drag.
Therefore, the final mass and the number do not depend on the initial stopping time.
Large planetesimals sweep small particles in the rotational direction.
Therefore the final mass depends on the size of the computational domain (paper II).

\subsection{Effect of Rotational Velocity Difference and Drag Law}
The difference in rotational velocity between dust and gas causes the radial migration of dust.
By neglecting the self-gravity and the time derivatives of Equation (\ref{eq:Hilleqx}) and (\ref{eq:Hilleqy}), we obtain the steady solution: \citep{Adachi1976, Weidenschilling1977}
\begin{equation}
  \tilde v_x = \frac{2 \tilde t_\mathrm{stop} \tilde v_\mathrm{dif}}{\tilde t_\mathrm{stop}^2+1}.
  \label{eq:driftvelo}
\end{equation}

\begin{figure}
  \begin{center}
	\plotone{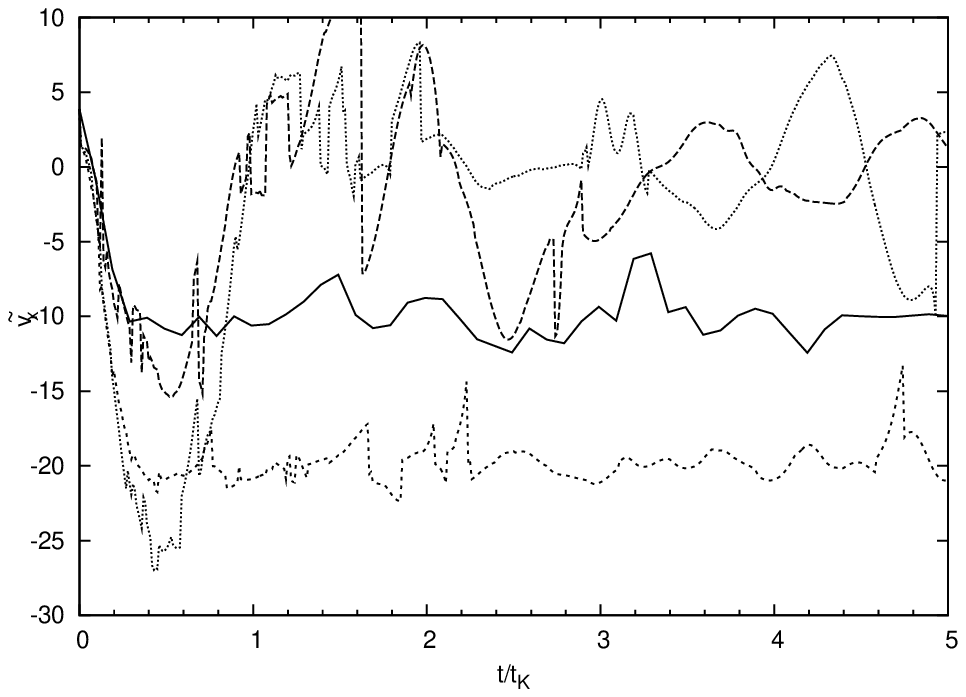}
  \end{center}
  \caption{The time evolution of the radial velocity $v_x$ of the largest planetesimal for models 100AC1 (\textit{solid curve}),
  100AS1 (\textit{dashed curve}), 100AC2 (\textit{short dashed curve}), and 100AS2 (\textit{dotted curve}). 
Note that the planetesimal whose velocity is plotted is not identical, i.e., the largest planetesimal changes as the other planetesimal grows.
Therefore the velocity changes abruptly then.}
  \label{fig:maxmass_vx.eps}
\end{figure}

Figure \ref{fig:maxmass_vx.eps} shows the time evolution of the radial velocity $\tilde v_x$ of the largest particle.
In these models, the initial stopping time is $\tilde t_\mathrm{stop,0}=1.0$.
The power-law index and the velocity difference are $(\alpha,\tilde v_\mathrm{dif}) = (0,10)$, $(2/3,10)$, $(0,20)$, and $(2/3,20)$ (models 100AC1, 100AS1, 100AC2, and 100AS2).
In all models, the radial velocity $\tilde v_x$ is negative at $t/t_\mathrm{K}=1 $.
This corresponds to the radially inward drift.
The rotational velocity of gas is slower than the Kepler velocity; dust experiences a headwind, which causes its inward migration.
Therefore their radial velocities become negative.
Although $\alpha$ is different, if $v_\mathrm{dif}$ is same, the time evolutions of $v_x$ are similar in the early phase.
This is because the planetesimals do not grow sufficiently in $t < 1 t_\mathrm{K}$.
If we estimate the terminal velocity from Equation (\ref{eq:driftvelo}) by the initial stopping time, they are the radial velocity $\tilde v_\mathrm{in}=10$ for $\tilde v_\mathrm{dif}=10$ and $\tilde v_\mathrm{in}=20$ for $\tilde v_\mathrm{dif}=20$.
In models for $\alpha=0$, the drag coefficient of a particle is independent of its size, thus the radial velocity converges to $\tilde v_x=10$ for $\tilde v_\mathrm{dif}=10$ and $\tilde v_x=20$ for $\tilde v_\mathrm{dif}=20$; the radial migration does not stop.
On the other hand, In models for $\alpha=2/3$, as planetesimals grow, the drag coefficient becomes small; the radial migration stops finally.
The radial velocity oscillates with the period of about $t_\mathrm{K}$.
This oscillation is caused by the epicycle motion.
The gas drag is so weak that the oscillation is not damped.

\section{Discussion}

\subsection{Comparison with Linear Analyses}

\cite{Youdin2005a} performed linear analyses of the dissipative
 gravitational instability of a dust layer.     
They studied the effect of the stopping time, the thickness of the dust
 layer, and the velocity dispersion.
We summarized the essence of their results in \S \ref{sec:disprel}.
The dust layer is always unstable for long wavelength modes.
Thus, we should compare the timescale of the gravitational instability
 with those of other processes. 
Under the thin disk approximation ($h=0$), we derived that simple condition that the gravitational instability dominates the
 other processes. 
Though the disk thickness cannot be neglected and the gravitational
 wakes is not axisymmetric, this condition corresponds to the occurrence
 of the gravitational wakes, as shown in \S \ref{sec:result}. 
To discuss the formation of the self-gravity wakes, Toomre's $Q$ value
 is more important than the Roche criterion $Q_{\rm R}$. 

They applied the growth rate of the gravitational instability to the gravitational collapse in the turbulent disks. 
They assumed the simple turbulence model and studied the particle response to the turbulence flow. 
The protoplanetary disk can be turbulent due to magnetorotational instability or Kelvin-Helmholtz instability.
They suggested that the standard hydrostatic relation
 $h \simeq \sqrt{2} \sigma_z / \Omega$ is not satisfied in
 the turbulent disk and the dust disk can be very thick.
The thin disk approximation is clearly not valid. 
The criterion of the formation of the self-gravitational wakes may
 change. 
We plan to study the effect of the turbulence in the next paper.

\subsection{Comparison with Hydrodynamic Simulations}

\citet{Yamoto2006} performed the two-dimensional numerical simulation of
 the gravitational instability of a dust layer.   
They found that the dust layer becomes extremely thin if the stopping
 time is long, such as $\tilde t_\mathrm{stop} > 0.1$.  
This is because the dust settling is faster than the growth of the
 gravitational instability. 
However, our results show that the gravitational instability occurs
 although the stopping time is long.  

There are some differences between our simulations and theirs. 
We treated the dust component as particles. 
But, \citet{Yamoto2006} treated it as a pressure-free fluid.
If the stopping time is short and the velocity dispersion is small,
 we can treat dust as a pressure-free fluid. 
The pressure-free fluid model becomes irrelevant for longer stopping times. 
For $\tilde t_\mathrm{stop} = 0.1$, we can use the pressure-free fluid, thus this should not cause the difference. 
In addition, the velocity dispersion stabilizes the gravitational instability generally. 
We neglected the dust back-reaction on the gas, but \citet{Yamoto2006} adopted two-component model and solved the interaction between them.
If the dust density is high and the stopping time is short, the dust
 back-reaction on the gas should be important. 
The resulting gas flow may change the criterion.
We performed three-dimensional local calculation. 
On the other hand, \citet{Yamoto2006} assumed that the the dust layer to be axisymmetric with respect to the rotational axis. 
The gravitational instability forms the non-axisymmetric structures such as the gravitational wakes \citep[e.g.,][]{Michikoshi2007, Wakita2008}. 
The motion in the azimuthal direction is important to study the structure of the gravitationally unstable disk.  

To understand the dynamics of planetesimal formation in gas, we need to perform three-dimensional two-component simulations. 
We should consider the dust back-reaction on the gas and solve the hydrodynamic equations consistently.
We will treat the dust component as particles because this method is applicable although the dust is loosely coupled to gas. 
We plan to study this point in the following papers of this series. 

\section{Summary}
We performed local $N$-body simulations of the planetesimal formation through gravitational instability.
In papers I and II, we adopted the gas-free model. 
However the gas is not negligible and is important to the formation of planetesimals.
As a first step to understand the effect of gas, we considered gas as the background laminar flow.
We neglected the dust back-reaction on the gas for the sake of simplicity. 
Laminar flow causes the dissipation of the kinetic energy and thus radial migration of dust.  

In \S 2, we summarized the results of the linear analysis of the dissipative gravitational instability. 
To handle the dispersion relation analytically, we imposed the assumptions: the isothermal equation of state, the infinitely thin disk, and the axisymmetric perturbation.
Simulations show that these assumptions are not valid, but they help us to understand the physical nature of the dissipative  gravitational instability.
We can use the growth rate to understand the simulation results.
The long wavelength modes are always secularly unstable although $Q>1$.
Therefore, we must compare with the timescale of the gravitational instability with other processes.
We provided the equation (\ref{eq:crit_Q}) for the critical $Q$ value. If $Q$ is smaller than the critical value, the gravitational instability is a dominant process. 

The numerical simulations show that the formation process of planetesimals is the same as that in the gas-free model.
The formation process is divided into three stages qualitatively: the formation of wake-like structures, the creation of planetesimal seeds, and their collisional growth. 
By the linear stability analysis, the dust layer in the laminar gas disk is secularly unstable although $Q>1$. 
The growth time of the dissipative gravitational instability is slower than the dust sedimentation and the decrease of the velocity dispersion when the initial $Q$ value is sufficiently large.
Thus, the disk shrinks vertically and $Q$ decreases.
As the velocity dispersion and the scale height decrease, the growth rate of the gravitational instability increases.
Finally, the gravitational instability becomes dominant.
Then wake-like structures are formed by the gravitational instability.  
These structures fragment into planetesimal seeds. 
Seeds grow rapidly owing to the mutual collisions.
Finally, almost all mass in the computational domain is absorbed by only one planetesimal in the calculation.

We investigated the dependence of results on the initial stopping time.
The $Q$ value decreases in the initial stage, and it reaches the minimum.
The decay time scale of $Q$ is proportional to the stopping time for $\tilde t_\mathrm{stop,0}<1$, which is the decay timescale of the velocity dispersion. 
The minimum $Q$ is determined by the balance between the dissipation and gravitational instability.
At the minimum Q, the wake-like structures are formed.
The time evolution of the ratio of the dust layer density to the Roche density $Q_R$ is similar to that of $Q$.
However the decay time scale of $Q_R$ is longer than that of $Q$.
The time when $Q_R=1$ or $Q_R=Q_{R,\mathrm{min}}$ does not correspond to the onset of the gravitational instability, such as the formation of the wake-like structures.
This indicates that the Roche criterion is not applicable in this system.
We investigated the time evolution of the number of planetesimal seeds.
We define a particle whose mass is larger than $M_\mathrm{linear}$ as a planetesimal seed, where $M_\mathrm{linear}$ is the planetesimal mass predicted by the linear theory $\pi \Sigma (\lambda_\mathrm{m}/2)^2$.
The number of planetesimals has a maximum value.
After the gravitational instability becomes a dominant process, planetesimal seeds form.
In this stage, the number of planetesimal seeds increases.
However, in the late stage, the number of seeds decreases by the mutual collisions among them.
The maximum of the number of the planetesimal seeds depends on the initial stopping time.
The maximum number of planetesimal seeds has a peak at $\tilde t_\mathrm{stop,0} \simeq 0.5$.
We are unable to identify a clear difference in the mass of the largest particle in the final state.
The final mass of the planetesimal depends on the size of the computational domain \citep{Michikoshi2009}.

We also investigated the models where the gas rotational velocity is slower than the Kepler velocity.
We confirmed that the planetesimal is able to form in the sub-Keplerian gas disk.
In the initial stage, particles experience a headwind, which causes its inward migration.
If we adopt Stokes' gas drag law, as planetesimals grow, the drag coefficient becomes small; the radial migration finally slows.

Almost all mass is absorbed by the largest planetesimal.
In the gas-free models, we showed that the final mass of planetesimals depends on the size of computational domain.
To investigate the final mass of planetesimals in detail, we should perform larger scale simulations.
This remains to be discussed further.
In addition, in this paper, we assumed gas to be laminar flow.
However, the gas may be turbulent. 
If the turbulence is strong, the particles are stirred up and the gravitational instability may be prevented.
The timescale for the dissipative gravitational instability is long
\citep{Ward1976,Youdin2005a}.  However, though the velocity dispersion of dust
particles is large and the timescale for the instability is long, the dust
layer may disrupt owing to the dissipative gravitational instability over time
in the turbulent disk.  In the next paper, we will investigate the particle
response to the turbulence and the gravitational instability in the turbulent
disk.

\acknowledgments{
Numerical simulations were carried out on the MUV system at Center for Computational Astrophysics, National Astronomical Observatory Japan.
This research was partially supported by MEXT (Ministry of Education, Culture,
Sports, Science and Technology), Japan, the Grant-in-Aid for Scientific
Research on Priority Areas, ``Development of Extra-Solar Planetary Science,''
and the Special Coordination Fund for Promoting Science and Technology, ``GRAPE-DR Project.''
S. I. is supported by Grants-in-Aid (16077202, and 18540238) from MEXT of Japan.
}

\end{document}